\documentclass[aps,prc,nofootinbib,groupedaddress,twocolumn,10pt,floatfix]{revtex4-2}

\usepackage{8li-trans}

\ifproofpre{}{\ifeof18
    \typeout{Version control information not updated!}
\else
    \immediate\write18{vc/vc}
\fi

\RequirePackage{xcolor}

\IfFileExists{vc/vc-info.tex}{\gdef\GITAbrHash{7316451}\gdef\VCRevision{\GITAbrHash}\gdef\VCDateText{August 27, 2022}\gdef\VCRevisionMod{\VCRevision}\gdef\VCRevisionMod{\VCRevision} }{
\gdef\GITAbrHash{unknown}\gdef\VCRevision{\GITAbrHash}\gdef\VCDateText{\today}\gdef\VCRevisionMod{\VCRevision}}
 }

\begin{document}

\title{\textit{Ab initio} estimation of $E2$ strengths in $\isotope[8]{Li}$ and its neighbors by normalization to the measured quadrupole moment}

\author{Mark A.~Caprio\,\orcidlink{0000-0001-5138-3740}}
\affiliation{Department of Physics and Astronomy, University of Notre Dame, Notre Dame, Indiana 46556-5670, USA}

\author{Patrick J.~Fasano\,\orcidlink{0000-0003-2457-4976}}
\affiliation{Department of Physics and Astronomy, University of Notre Dame, Notre Dame, Indiana 46556-5670, USA}

\date{\ifproofpre{\today}{\VCDateText}}

\begin{abstract}
  \relax
For electric quadrupole ($E2$) observables, which depend on the large-distance
tails of the nuclear wave function, \textit{ab initio} no-core configuration
interaction (NCCI) calculations converge slowly, making meaningful predictions
challenging to obtain.
Nonetheless, the calculated values for different $E2$ matrix elements,
particularly those involving levels with closely-related structure
(\textit{e.g.}, within the same rotational band) are found to be robustly
proportional.  This observation suggests that a known value for one observable
may be used to determine the overall scale of $E2$ strengths, and thereby
provide predictions for others.
In particular, we demonstrate that meaningful predictions for $E2$ transitions
may be obtained by calibration to the ground-state quadrupole moment.
We test this approach for well-measured low-lying $E2$ transitions in
$\isotope[7]{Li}$ and $\isotope[9]{Be}$, then provide predictions for
transitions in $\isotope[8]{Li}$ and $\isotope[9]{Li}$.  In particular, we address the
$2^+\rightarrow1^+$ transition in $\isotope[8]{Li}$, for which the reported
measured strength exceeds \textit{ab initio} Green's function Monte Carlo (GFMC)
predictions by over an order of magnitude.
\relax
 \end{abstract}

\ifproofpre{}{\preprint{Git hash: \VCRevisionMod}}

\maketitle

\section{Introduction}
Electric quadrupole ($E2$) observables provide key measures of nuclear
collective
structure~\cite{bohr1998:v2,casten2000:ns,rowe2010:collective-motion}, in
particular, rotation and deformation.  However, \textit{ab initio} calculations
for $E2$ observables are notoriously challenging to
obtain~\cite{bogner2008:ncsm-converg-2N,maris2013:ncsm-pshell,odell2016:ir-extrap-quadrupole}.
Since $E2$ observables are sensitive to the large-distance tails of the nuclear
wave function, they are slowly convergent in \textit{ab initio} no-core
configuration interaction (NCCI), or no-core shell model (NCSM),
approaches~\cite{barrett2013:ncsm}, which conventionally rely upon an
oscillator-basis expansion of the wave function.  In practical calculations, the
basis for the many-body space must be truncated to finite size.  The results can
therefore, at best, only approximate the $E2$ predictions which would be
obtained by solving the full (untruncated) many-body problem for a given
internucleon interaction.  While one may attempt to improve the many-body
calculation by various means (\textit{e.g.},
Refs.~\cite{roth2007:it-ncsm-40ca,dytrych2008:sp-ncsm,*dytrych2016:su3ncsm-12c-efficacy,vorabbi2019:7be-7li-ncsmc,mccoy2020:spfamilies,fasano2022:natorb})
so as to improve convergence of $E2$ observables, the accuracy is nonetheless
severely limited by computational constraints.

We may thus, alternatively, seek indirect ways to circumvent the convergence
challenges affecting $E2$ observables.  In particular, the convergence patterns of calculated $E2$ matrix
elements are often strongly
correlated~\cite{caprio2013:berotor,*maris2015:berotor2,*maris2019:berotor2-ERRATUM,calci2016:observable-correlations-chiral,henderson2019:7be-coulex,caprio2019:bebands-sdanca19,caprio2020:bebands,caprio2021:emratio},
especially for matrix elements involving states with similar structure.  This
suggests~\cite{calci2016:observable-correlations-chiral} that, if one $E2$
matrix element is well known from experiment (or, in principle, a complementary
\textit{ab initio} calculation using an alternative many-body method), a meaningful prediction may then be made for another, correlated
$E2$ matrix element.  Calci and
Roth~\cite{calci2016:observable-correlations-chiral} use the well-measured $E2$
strength between the ground state and first excited state, in $\isotope[6]{Li}$
and $\isotope[12]{C}$, to obtain a prediction for the elusive excited-state
quadrupole moment.

\begin{figure}
\begin{center}
\includegraphics[width=\ifproofpre{1}{1}\hsize]{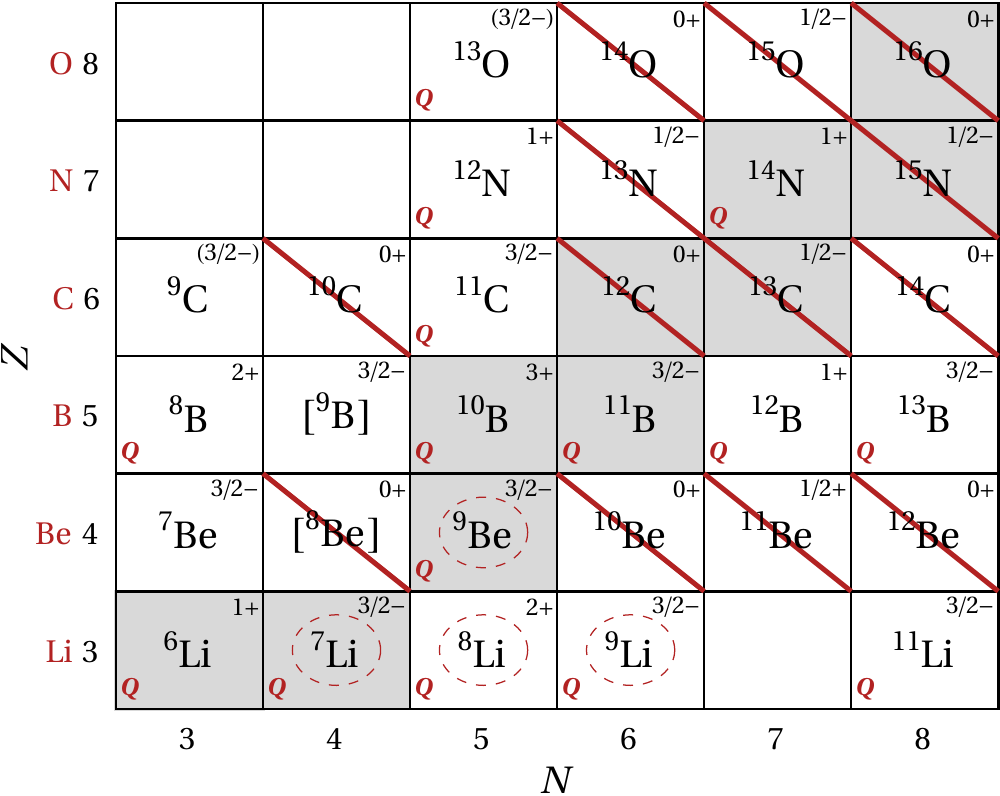}
\end{center}
\caption{Nuclides with measured ground-state quadrupole
  moments~\cite{stone2016:e2-moments} (indicated with the letter ``$Q$'') in the
  $p$ shell.  Particle-bound nuclides are designated by name, while brackets
  indicate a particle-unbound but narrow ($\lesssim 1\,\keV$) ground-state
  resonance, and shading indicates stable nuclides. The ground-state angular
  momentum and parity are
  given~\cite{npa2002:005-007,npa2004:008-010,npa2012:011,npa2017:012,npa1991:013-015}
  (upper right), while slashes serve to exclude those nuclei (with $J\leq1/2$)
  for which the ground-state angular momentum does not support a quadrupole
  moment.  The nuclide $\isotope[8]{Li}$ and its neighbors considered in this
  work are highlighted (dashed circles).  Figure adapted from
  Ref.~\cite{caprio2021:emratio}.
}
\label{fig:nuclear-chart}
\end{figure}

Conversely, in the present work, we demonstrate the viability of the
ground-state quadrupole moment as a calibration reference by which to generate
predictions of $E2$ strengths, through robust \textit{ab initio} NCCI
predictions of the dimensionless ratio $B(E2)/(eQ)^2$, in which systematic
truncations errors in the calculated $E2$ matrix elements cancel.  The
ground-state quadrupole moment is well measured for many
nuclei~\cite{stone2016:e2-moments}, as summarized for $p$-shell nuclides in
Fig.~\ref{fig:nuclear-chart}.  Calibration to this observable is subject to the
fundamental constraint that the ground state angular momentum must admit a
nonvanishing quadrupole moment ($J\geq1$), as well as practical constraints that
measurement must be feasible~\cite{neugart2006:nuclear-moments}, including that
the ground state must be particle bound.

The case of $\isotope[8]{Li}$ is of particular interest, as an instance in which
this approach may be applied to obtain \textit{ab initio} insight, given the
anomalously enhanced strength reported for the transition between the $2^+$
ground state and $1^+$ first excited state of this nuclide.  This $E2$ strength
has been measured through Coulomb excitation of $\isotope[8]{Li}$ in a
radioactive beam experiment, yielding $B(E2;2^+\rightarrow
1^+)=55(15)\,e^2\fm^4$~\cite{brown1991:8li-coulex,npa2004:008-010}, or, in terms
of the Weisskopf single-particle estimate~\cite{weisskopf1951:estimate},
$\approx58\,\Wu$ (The gamma decay lifetime of the $1^+$ state instead yields
only information on the $M1$ strength~\cite{npa2004:008-010}.)  This is among
the most enhanced $E2$ transition strengths reported in a $p$-shell
nuclide~\cite{npa2002:005-007,npa2004:008-010,npa2012:011,npa2017:012,npa1991:013-015}.
Compare, \textit{e.g.}, $B(E2; 3/2^-\rightarrow1/2^-)\approx 10\,\Wu$ for the
analogous (upward) transition from the ground state of neighboring
$\isotope[7]{Li}$~\cite{npa2002:005-007}, or $B(E2;
3/2^-\rightarrow5/2^-)\approx 42\,\Wu$ similarly in neighboring
$\isotope[9]{Be}$~\cite{npa1991:013-015}.

However, Green's function Monte Carlo (GFMC)
calculations~\cite{pastore2013:qmc-em-alt9} give a predicted strength nearly two
orders of magnitude smaller, at
$0.83(7)\,e^2\fm^4$~\cite{pastore2013:qmc-em-alt9}.  Moreover, we note that such
enhancement in $\isotope[8]{Li}$ would be particularly remarkable, given that it
cannot be explained in terms of in-band rotational collectivity, while the
aforementioned transitions in neighboring $\isotope[7]{Li}$ and
$\isotope[9]{Be}$ are ostensibly rotational in nature~\cite{caprio2020:bebands}.
Even if the $2^+$ ground state is taken to be a $K=2$ rotational band head, this band would
have no $J=1$ member.

We first establish the expected form for the correlation between $B(E2)$ and
quadrupole moment observables, through the dimensionless ratio $B(E2)/(eQ)^2$
(Sec.~\ref{sec:methods}), and demonstrate the robust convergence of this ratio
for experimentally well-measured $E2$ transition strengths, between the ground
state and first excited state (of the same parity), in $\isotope[7]{Li}$ and
$\isotope[9]{Be}$ (Sec.~\ref{sec:results-benchmark}).  We then return to the
anomalous $2^+\rightarrow1^+$ transition in $\isotope[8]{Li}$ and other
unmeasured $E2$ strengths to low-lying states in $\isotope[8]{Li}$ and
$\isotope[9]{Li}$ (Sec.~\ref{sec:results-prediction}).

 \section{Dimensionless ratio}
\label{sec:methods}

The $E2$ reduced transition probability depends upon the square of a reduced
matrix element of the $E2$ operator, as
\begin{equation}
  \label{eqn:be2}
  B(E2;J_i\rightarrow J_f)
  \propto \abs{\trme{J_f}{
      \,Q_2\,
    }{J_i}}^2,
\end{equation}
while the quadrupole moment, originally defined in terms of the stretched matrix element
\begin{math}
  \label{eqn:q}
  \tme
      {JJ}{
        \,Q_{2,0}\,
      }{JJ},
\end{math}
is simply proportional to a reduced matrix element,
as
\begin{equation}
  eQ(J)
  \propto\trme{J}{
    \,Q_2\,
  }{J}.
\end{equation}
The sensitivity of each observable to the large-distance
properties of the nuclear wave function arises from the $r^2$ dependence of the
$E2$ operator~\cite{suhonen2007:nucleons-nucleus},
\begin{math}
  Q_{2\mu}=\sum_{i\in p}e
  r_{i}^2Y_{2\mu}(\uvec{r}_{i}),
\end{math}
where the summation runs over the (charged) protons.  The ratio
\begin{equation}
  \label{eqn:ratio}
  \frac{B(E2)}{(eQ)^2}\propto
  \Biggl\lvert
  \frac{
\trme{J_f}{Q_2}{J_i}
  }{
\trme{J}{Q_2}{J}
  }
  \Biggr\rvert^2
\end{equation}
is dimensionless, and involves like powers of reduced matrix elements of the
$E2$ operator in the numerator and denominator.  We thus have reason to hope for
at least partial cancellation of the error arising in these matrix elements due
to truncation of the nuclear wave functions.

\section{Illustration for $\isotope[7]{Li}$ and $\isotope[9]{Be}$}
\label{sec:results-benchmark}

In the NCCI approach, the true results of solving the many-body problem in the
full many-body space would be obtained if the full, infinite oscillator basis
could be used.  However, for finite calculations, results depend upon the
subspace spanned by the truncated basis.  Thus they depend both
upon the maximum number~$\Nmax$ of oscillator excitations allowed within the
configurations making up the many-body basis, and upon the oscillator length of
the underlying single-particle states (or, equivalently, the oscillator
parameter $\hw$~\cite{suhonen2007:nucleons-nucleus}).  Convergence is recognized
when the calculated results become insensitive to increases in $\Nmax$ and to
variation in $\hw$ (see, \textit{e.g.},
Refs.~\cite{bogner2008:ncsm-converg-2N,maris2013:ncsm-pshell,caprio2020:bebands}).

\begin{figure*}
\centering
\includegraphics[width=\ifproofpre{1.0}{1.0}\hsize]{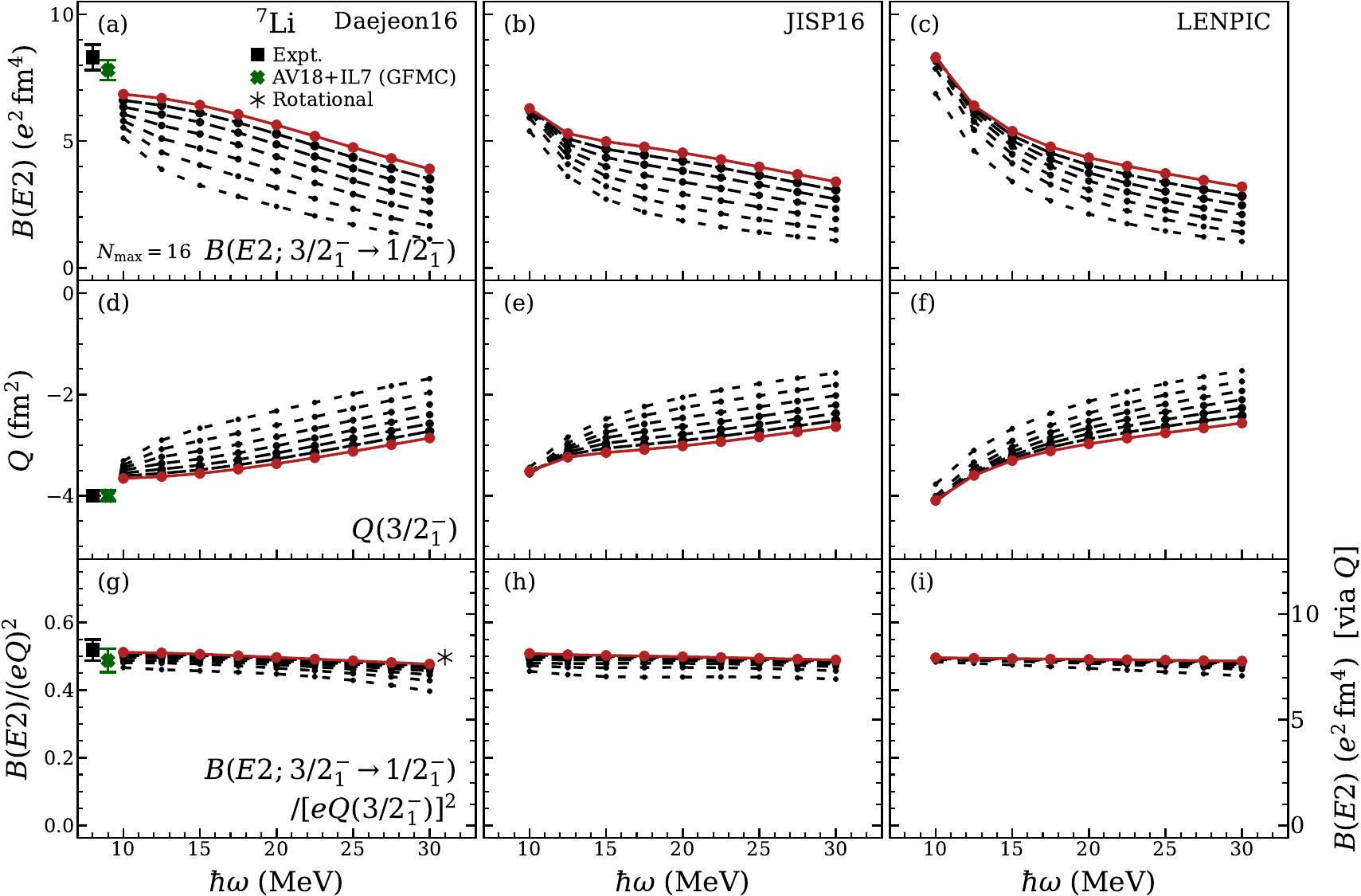}
\caption{Convergence of \textit{ab initio} NCCI calculated observables for
  $\isotope[7]{Li}$: (top)~the $3/2^-\rightarrow1/2^-$ $E2$ strength,
  (middle)~the electric quadrupole moment of the $3/2^-$ ground state, and
  (bottom)~the dimensionless ratio $B(E2)/(eQ)^2$ constructed from the preceding
  two observables.  Results are shown for the (left)~Daejeon16, (center)~JISP16,
  and (right)~LENPIC interactions. When calibrated to the experimental
  quadrupole moment, the ratio provides a prediction for the absolute $B(E2)$
  (scale at right).  Calculated values are shown as functions of the basis
  parameter $\hw$, for successive even values of $\Nmax$ (increasing symbol size
  and longer dashing), from $\Nmax=4$ (short dashed curves) to $16$ (solid
  curves).  For comparison, experimental
  values~\cite{stone2016:e2-moments,npa2002:005-007} (squares), GFMC AV18+IL7
  predictions~\cite{pastore2013:qmc-em-alt9} (crosses), and the rotational ratio
  (asterisk) are also shown.  }
\label{fig:be2-norm-qp-scan-7li}
\end{figure*}

Let us first consider the convergence of the calculated $3/2^-\rightarrow 1/2^-$
$E2$ strength for $\isotope[7]{Li}$, shown in
Fig.~\ref{fig:be2-norm-qp-scan-7li}(a), as obtained using the Daejeon16
internucleon interaction~\cite{shirokov2016:nn-daejeon16}.  This interaction is
based on the two-body part of the Entem-Machleidt \nthreelo{} chiral effective
field theory ($\chi$EFT) interaction~\cite{entem2003:chiral-nn-potl}, softened
via a similarity renormalization group (SRG)
transformation~\cite{bogner2007:srg-nucleon} so as to provide comparatively
rapid convergence, and then adjusted via a phase-shift equivalent transformation
to better describe nuclei with $A\leq16$ while still maintaining rapid
convergence.  Calculations are carried out using the NCCI code
MFDn~\cite{aktulga2014:mfdn-scalability,shao2018:ncci-preconditioned,cook2022:mfdn-directives-waccpd21}.
(Comprehensive plots and tabulations of calculated observables, as
functions of $\Nmax$ and $\hw$, are provided in the Supplemental
Material~\cite{supplemental-material}.)

The values along each curve in Fig.~\ref{fig:be2-norm-qp-scan-7li}(a) represent the
results of calculations carried out with the same basis truncation $\Nmax$ (from
short dashes for $\Nmax=4$ to solid lines for $\Nmax=16$) and differing $\hw$.
While there is perhaps some tendency towards flattening of these curves with
respect to $\hw$ (``shouldering'') and compression of successive curves with
respect to $\Nmax$, the calculated values are still steadily increasing with
increasing $\Nmax$.  At best, we might crudely estimate the true value which
would be obtained for the given internucleon interaction in the full,
untruncated many-body space.

A similar convergence pattern is found for the calculated $3/2^-$ ground state
quadrupole moment [Fig.~\ref{fig:be2-norm-qp-scan-7li}(d)], where, however, the
curves are inverted due to the negative sign on the quadrupole moment.  (For
further discussion of the convergence of this and other quadrupole moments in
NCCI calculations, see Ref.~\cite{caprio2021:emratio}.)  With each increment in
$\Nmax$, the relative (fractional) change between calculated values of the
quadrupole moment is smaller than for the $B(E2)$.  This is to be expected, as
the quadrupole moment is simply proportional to a matrix element of the $E2$
operator, while the $B(E2)$ is proportional to the square of such a matrix
element, and (as in elementary error analysis) squaring a quantity doubles
relative changes in that quantity.  However, one may again at best attempt a
crude estimate of the value which would be obtained in the full, untruncated
many-body space.

In $\isotope[7]{Li}$, both the $E2$ strength and the quadrupole moment are known
experimentally, with measured values of $B(E2; 3/2^-\rightarrow
1/2^-)=8.3(5)\,e^2\fm^4$~\cite{npa2002:005-007} and
$Q(3/2^-)=-4.00(3)\,\fm^2$~\cite{stone2016:e2-moments,npa2002:005-007} (squares
in Fig.~\ref{fig:be2-norm-qp-scan-7li}).  While the NCCI calculated values for
both the $B(E2)$ [Fig.~\ref{fig:be2-norm-qp-scan-7li}(a)] and quadrupole moment
[Fig.~\ref{fig:be2-norm-qp-scan-7li}(d)] are increasing in the general direction
of the experimental result, these poorly-converged results do not permit
meaningful, quantitative comparison.

However, let us now take the dimensionless ratio of the form defined
in~(\ref{eqn:ratio}) for these observables, namely, $B(E2;3/2^-\rightarrow
1/2^-)/[eQ(3/2^-)]^2$, with the result shown in
Fig.~\ref{fig:be2-norm-qp-scan-7li}(g).  We find a near complete elimination of
the $\hw$ dependence, at the higher $\Nmax$ shown, as well as a radical
compression of the curves for successive $\Nmax$.  Calibrating to the known
ground-state quadrupole moment~\cite{stone2016:e2-moments} gives the scale shown at far right
[Fig.~\ref{fig:be2-norm-qp-scan-7li} (bottom)].  An estimated ratio of
$B(E2;3/2^-\rightarrow 1/2^-)/[eQ(3/2^-)]^2\approx0.50$ yields
$B(E2;3/2^-\rightarrow 1/2^-)\approx8\,e^2\fm^4$.  The predicted ratio
$B(E2)/(eQ)^2$ is consistent, to within uncertainties, with the experimental
ratio of $0.52(3)$, and the resulting $B(E2)$ is similarly within uncertainties
of the experimental strength.

\begin{figure}
\centering
\includegraphics[width=\ifproofpre{1.0}{0.5}\hsize]{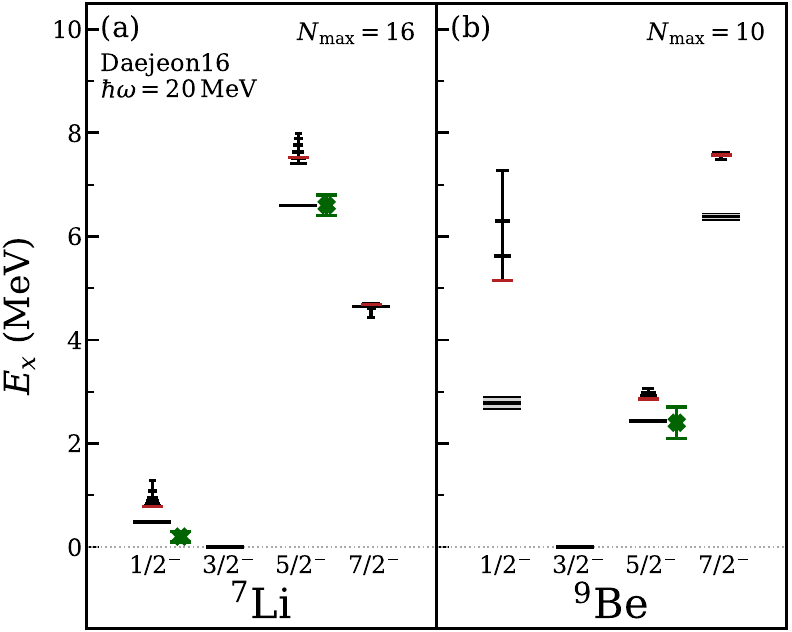}
\caption{Calculated excitation energies of low-lying states in
  (a)~$\isotope[7]{Li}$ and (b)~$\isotope[9]{Be}$, with angular momentum and
  parity as indicated at bottom, as obtained with the Daejeon16 interaction.
  Calculated values are shown at fixed $\hw=20\,\MeV$ and varying $\Nmax$
  (increasing symbol size), from $\Nmax=4$ to the maximum value indicated (at
  top).  Experimental energies~\cite{npa2002:005-007,npa2004:008-010} are shown
  (horizontal line and error band) where available, as are the GFMC AV18+IL7
  predictions~\cite{pastore2013:qmc-em-alt9} (crosses) (see Table~III of
  Ref.~\cite{carlson2015:qmc-nuclear}).  }
\label{fig:excitation-teardrop-benchmark}
\end{figure}

From a physical viewpoint, the close-lying $3/2^-$ ground state and $1/2^-$
excited state in $\isotope[7]{Li}$ are interpreted as members of a $K=1/2$
rotational
band~\cite{millener2001:light-nuclei,*millener2007:p-shell-hypernuclei}, where
the energy order is inverted due to Coriolis
staggering~\cite{rowe2010:collective-motion}.  For context, the calculated and
experimental excitation energies of the yrast levels are shown in
Fig.~\ref{fig:excitation-teardrop-benchmark}(a) (see also Fig.~3 of
Ref.~\cite{caprio2020:bebands} and Fig.~2 of Ref.~\cite{mccoy2020:spfamilies}
for more extensive calculated level schemes of the mirror nuclide
$\isotope[7]{Be}$, obtained with the same Daejeon16 interaction).  The
rotational model yields $B(E2;3/2_{K=1/2}\rightarrow
1/2_{K=1/2})/[eQ(3/2_{K=1/2})]^2\approx0.497$, indicated by the asterisk in
Fig.~\ref{fig:be2-norm-qp-scan-7li}(g).  We are thus seeing close consistency
between \textit{ab initio} theory and experiment, both of which are
well explained by a simple rotational
picture~\cite{caprio2013:berotor,*maris2015:berotor2}.

To explore the dependence upon internucleon interaction, let us consider the
results for these same observables, but from calculations based on the JISP16
[Fig.~\ref{fig:be2-norm-qp-scan-7li} (center)] and LENPIC
[Fig.~\ref{fig:be2-norm-qp-scan-7li} (right)] internucleon interactions.  The
  phenomenological JISP16 interaction~\cite{shirokov2007:nn-jisp16} is obtained
  by $J$-matrix inverse scattering from nucleon-nucleon scattering data, and,
  like Daejeon16, adjusted via a phase-shift equivalent transformation to better
  describe nuclei with $A\leq16$.  The LENPIC
  interaction~\cite{epelbaum2015:lenpic-n4lo-scs,epelbaum2015:lenpic-n3lo-scs}
  is a modern chiral EFT interaction (we specifically take the two-body part, at
  \ntwolo{}, with a semi-local coordinate-space regulator of length scale
  $R=1\,\fm$, and, for purposes of illustration, use the bare interaction with
  no SRG transformation).

For the $B(E2)$ itself, there is at best minimal suggestion of convergence, or
shouldering, in the JISP16 results [Fig.~\ref{fig:be2-norm-qp-scan-7li}(b)], and
essentially no sign of convergence in the LENPIC results
[Fig.~\ref{fig:be2-norm-qp-scan-7li}(c)].  The same may be said for the computed
quadrupole moments [Fig.~\ref{fig:be2-norm-qp-scan-7li}(e,f)].  Nonetheless,
taking the dimensionless ratio $B(E2)/(eQ)^2$
[Fig.~\ref{fig:be2-norm-qp-scan-7li}(h,i)] again leads to a rapidly convergent
quantity, from which the $\hw$ dependence has largely been eliminated, and the
changes with successive $\Nmax$ rapidly decrease.  The resulting values for the
ratio, as obtained with these interactions, are closely consistent both with that
obtained from the Daejeon16 interaction [Fig.~\ref{fig:be2-norm-qp-scan-7li}(g)]
and with experiment.

Predictions for this same quadrupole moment and transition matrix element in
$\isotope[7]{Li}$ have previously been reported~\cite{pastore2013:qmc-em-alt9}
from \textit{ab initio} Green's function Monte Carlo
(GFMC)~\cite{carlson2015:qmc-nuclear} calculations, based on the Argonne
$v_{18}$ (AV18) two-nucleon~\cite{wiringa1995:nn-av18} and Illinois-7 (IL7)
three-nucleon~\cite{pieper2001:3n-il2} potentials.  These predictions, shown as
crosses in Fig.~\ref{fig:be2-norm-qp-scan-7li} (left), are subject to Monte Carlo
statistical errors, so the calculational uncertainties are of a qualitatively
different nature from those entering into the NCCI calculations.  In particular,
the GFMC calculated values for the $E2$ transition strength
[Fig.~\ref{fig:be2-norm-qp-scan-7li}(a)] and quadrupole moment
[Fig.~\ref{fig:be2-norm-qp-scan-7li}(d)] may meaningfully be compared directly
with experiment, without taking a ratio to cancel truncation errors, and we see
agreement within uncertainties in both cases.  Nonetheless, for comparison with
the NCCI results, we may recast these GFMC results as a ratio $B(E2)/(eQ)^2$
[cross in Fig.~\ref{fig:be2-norm-qp-scan-7li}(g)], where we find consistency
with experiment (again), but now also with the NCCI predictions for the ratio.

\begin{figure*}
\centering
\includegraphics[width=\ifproofpre{0.75}{0.95}\hsize]{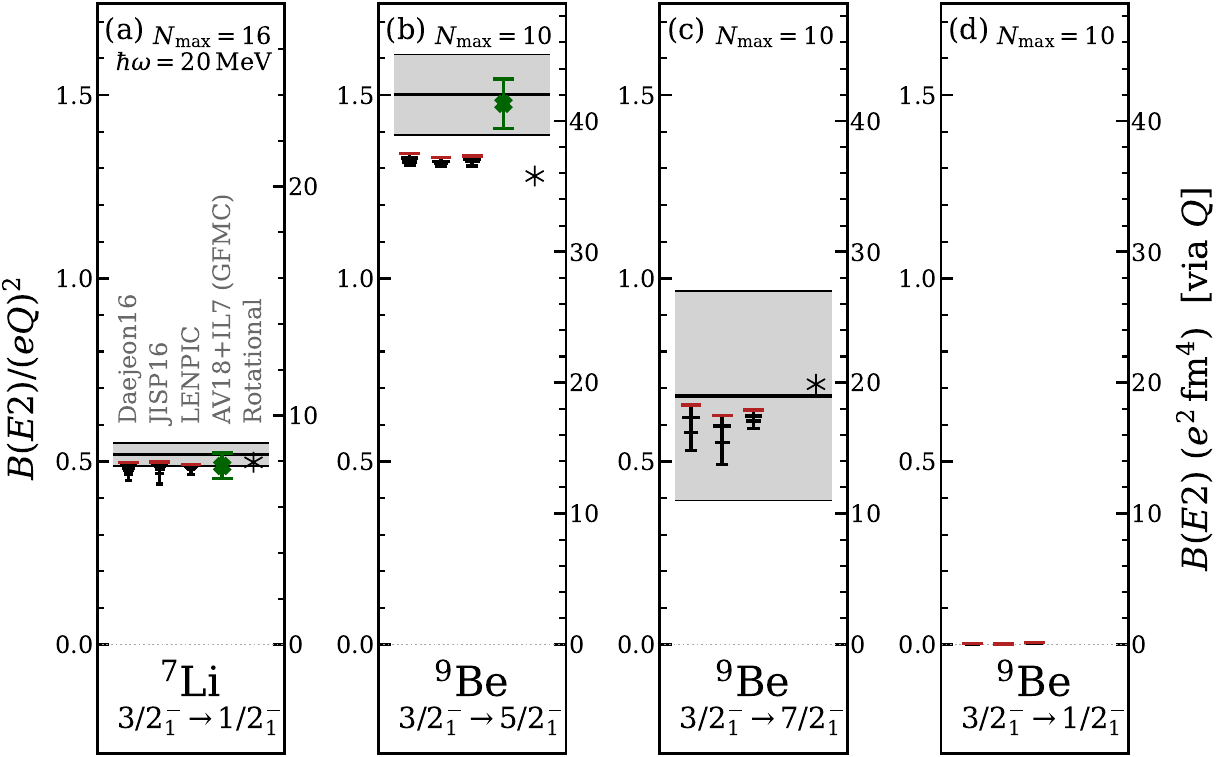}
\caption{Calculated ratios of the form $B(E2)/(eQ)^2$, for excitation to low-lying states
  in $\isotope[7]{Li}$ and $\isotope[9]{Be}$, obtained with the Daejeon16, JISP16,
  and LENPIC interactions (from left to right, for each transition).
  Calculated values are shown at fixed
  $\hw=20\,\MeV$ and varying $\Nmax$ (increasing
  symbol size), from $\Nmax=4$ to the maximum value indicated (at top).
  When calibrated to the experimental quadrupole moment~\cite{stone2016:e2-moments}, this
  ratio provides an estimate for the absolute $B(E2)$ (scale at right).
  Experimental
  results~\cite{npa2002:005-007,npa2004:008-010,stone2016:e2-moments} are shown
  (horizontal line and error band) where available, as are the GFMC AV18+IL7
  predictions~\cite{pastore2013:qmc-em-alt9} (crosses) and rotational ratios
  (asterisks).
}
\label{fig:be2-teardrop-benchmark}
\end{figure*}

To provide for convenient comparison across calculations and (in the following
discussion) transitions, we take a ``slice'' through these NCCI results in
Fig.~\ref{fig:be2-teardrop-benchmark}(a), which shows convergence with $\Nmax$
at fixed $\hw$ (chosen as $\hw=20\,\MeV$, based on the approximate location of
the variational energy minimum for the ground state, although this location
varies somewhat by nuclide and interaction).  We may again readily compare the
NCCI results with experiment (horizontal lines and shaded error bands), GFMC
AV18+IL7 predictions (crosses), and the rotational model (asterisks), where
applicable.

In $\isotope[9]{Be}$, the $E2$ transition from the $3/2^-$ ground state to the
$5/2^-$ excited state (a narrow resonance just above the neutron threshold, with
a width of $\approx0.8\,\keV$~\cite{npa2002:005-007}) is interpreted as an
in-band transition within the ground-state ($K=3/2$) rotational
band~\cite{millener2001:light-nuclei,*millener2007:p-shell-hypernuclei}.  For
context, calculated and experimental excitation energies of the
(normal-parity~\cite{lane1960:reduced-widths}) yrast levels of
$\isotope[9]{Be}$, including the $J=3/2$, $5/2$, and $7/2$ members of the ground
state $K=3/2$ band and the excited $K=1/2$ band head, are shown in
Fig.~\ref{fig:excitation-teardrop-benchmark}(b) (see also Fig.~1 of
Ref.~\cite{caprio2020:bebands} for a more extensive calculated level scheme,
obtained with the same Daejeon16 interaction).

\begin{figure}
\centering
\includegraphics[width=\ifproofpre{1.0}{0.5}\hsize]{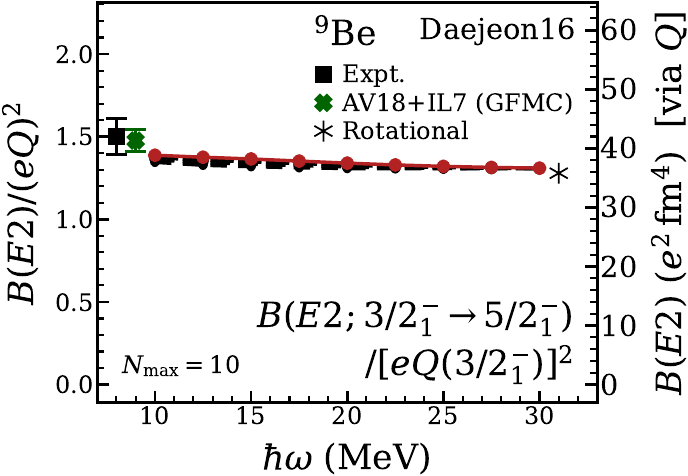}
\caption{Convergence of the \textit{ab initio} NCCI calculated dimensionless ratio
  $B(E2)/(eQ)^2$, for $\isotope[9]{Be}$, constructed from the $3/2^-\rightarrow5/2^-$ $E2$ strength and the
  electric quadrupole moment of the $3/2^-$ ground state.
  Results are shown for the Daejeon16 interaction. When calibrated to the experimental quadrupole
  moment, the ratio provides a prediction for the absolute $B(E2)$ (scale at
  right).  Calculated values are shown as functions of the basis parameter
  $\hw$, for successive even values of $\Nmax$ (increasing symbol size and longer
  dashing), from $\Nmax=4$ (short dashed curves) to $10$ (solid curves).  For
  comparison, the experimental ratio~\cite{stone2016:e2-moments,npa2004:008-010}
  (square), GFMC AV18+IL7 prediction~\cite{pastore2013:qmc-em-alt9} (cross), and
  rotational ratio (asterisk) are also shown.  }
\label{fig:be2-norm-qp-scan-9be-daejeon16-ratio}
\end{figure}

The dimensionless ratio $B(E2)/(eQ)^2$, as obtained with the Daejeon16
interaction, is shown in Fig.~\ref{fig:be2-norm-qp-scan-9be-daejeon16-ratio}, and
similar results are obtained with the other two interactions considered above,
as summarized in Fig.~\ref{fig:be2-teardrop-benchmark}(b).  Again, taking the
dimensionless ratio largely eliminates the $\hw$ dependence of the results and
yields rapid convergence with respect to $\Nmax$.  Calibrating to the known
ground-state quadrupole moment~\cite{stone2016:e2-moments} gives the scale shown
at right.  An estimated ratio of
$B(E2;3/2^-\rightarrow5/2^-)/[eQ(3/2^-)]^2\approx1.3\text{--}1.4$ yields
$B(E2;3/2^-\rightarrow 5/2^-)\approx36\text{--}39\,e^2\fm^4$.  The NCCI results
for this ratio (with all three interactions) lie just below the uncertainty
ranges for the experimental ratio (square) and for the GFMC AV18+IL7 predictions
(cross), and just above the ratio of $B(E2;3/2_{K=3/2}\rightarrow 5/2_{K=3/2})/[eQ(3/2_{K=3/2})]^2\approx1.279$ for an ideal rotational
description (asterisk).

For the in-band transition to the $7/2^-$ band member in $\isotope[9]{Be}$
[Fig.~\ref{fig:be2-teardrop-benchmark}(c)], the \textit{ab initio} predictions
for the ratio $B(E2;3/2^-\rightarrow7/2^-)/[eQ(3/2^-)]^2$ are consistent across
choice of interaction and closely agree with the rotational value
$B(E2;3/2_{K=3/2}\rightarrow 7/2_{K=3/2})/[eQ(3/2_{K=3/2})]^2\linebreak[0]\approx0.711$,
while lying well within generous experimental uncertainties.

The strength of any interband $E2$ transition to the $1/2^-$ band head is
experimentally unknown~\cite{npa2004:008-010}.  However, the present NCCI
calculations give a ratio $B(E2)/(eQ)^2$ which is essentially vanishing on the
scale of Fig.~\ref{fig:be2-teardrop-benchmark}(d).  The calculated ratios
$B(E2;3/2^-\rightarrow1/2^-)/[eQ(3/2^-)]^2\lesssim0.005$ suggest
$B(E2;3/2^-\rightarrow1/2^-)\lesssim0.2\,e^2\fm^4$.  In a rotational
description, the interband $E2$ strength depends upon the interband intrinsic
$E2$ matrix element~\cite{bohr1998:v2,casten2000:ns,rowe2010:collective-motion}, and
a limit on the ratio $B(E2)/(eQ)^2$ may be translated, through
appropriate Clebsch-Gordan factors, into a limit on the ratio of in-band and
interband intrinsic matrix elements.

\section{Predictions for $\isotope[8]{Li}$ and $\isotope[9]{Li}$}
\label{sec:results-prediction}

\begin{figure*}
\centering
\includegraphics[width=\ifproofpre{1.0}{1.0}\hsize]{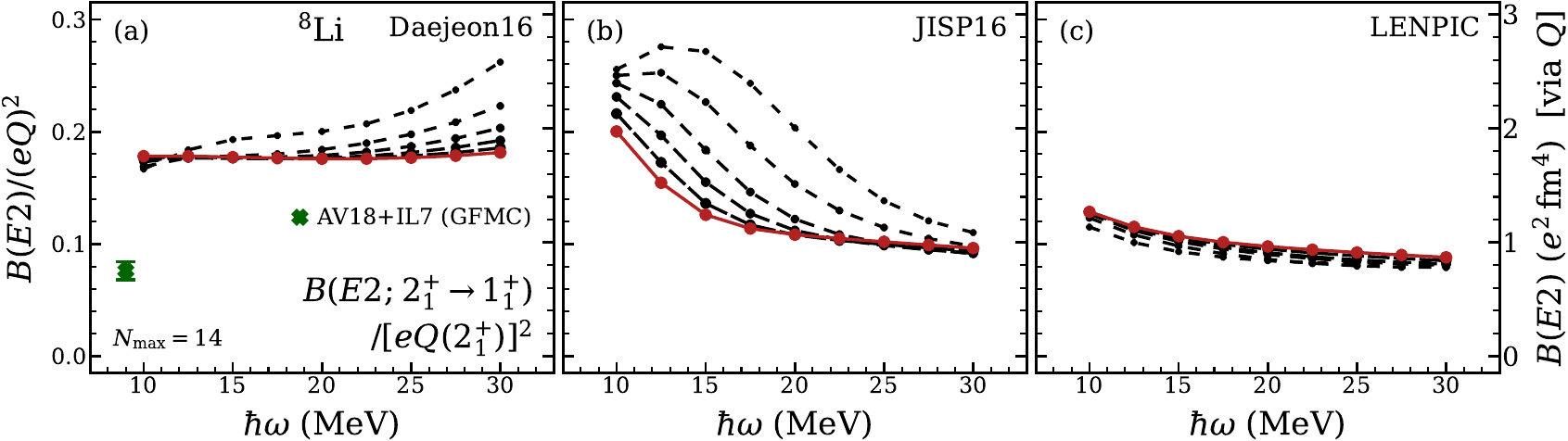}
\caption{Convergence of the \textit{ab initio} NCCI calculated dimensionless ratio
  $B(E2)/(eQ)^2$, for $\isotope[8]{Li}$, constructed from the
  $2^+\rightarrow1^+$ $E2$ strength and the electric quadrupole moment of the
  $2^+$ ground state.  Results are shown for the (a)~Daejeon16, (b)~JISP16, and
  (c)~LENPIC interactions. When calibrated to the experimental quadrupole
  moment, the ratio provides a prediction for the absolute $B(E2)$ (scale at
  right).  Calculated values are shown as functions of the basis parameter
  $\hw$, for successive even values of $\Nmax$ (increasing symbol size and longer
  dashing), from $\Nmax=4$ (short dashed curves) to $14$ (solid curves).  For
  comparison, the GFMC AV18+IL7 prediction~\cite{pastore2013:qmc-em-alt9}
  (cross) is shown, while the experimental
  ratio~\cite{stone2016:e2-moments,npa2004:008-010}, corresponding to the reported
  $E2$ strength of $55(15)\,e^2\fm^4$~\cite{brown1991:8li-coulex}, lies off
  scale.  }
\label{fig:be2-norm-qp-scan-8li}
\end{figure*}
\begin{figure}
\centering
\includegraphics[width=\ifproofpre{1.0}{0.5}\hsize]{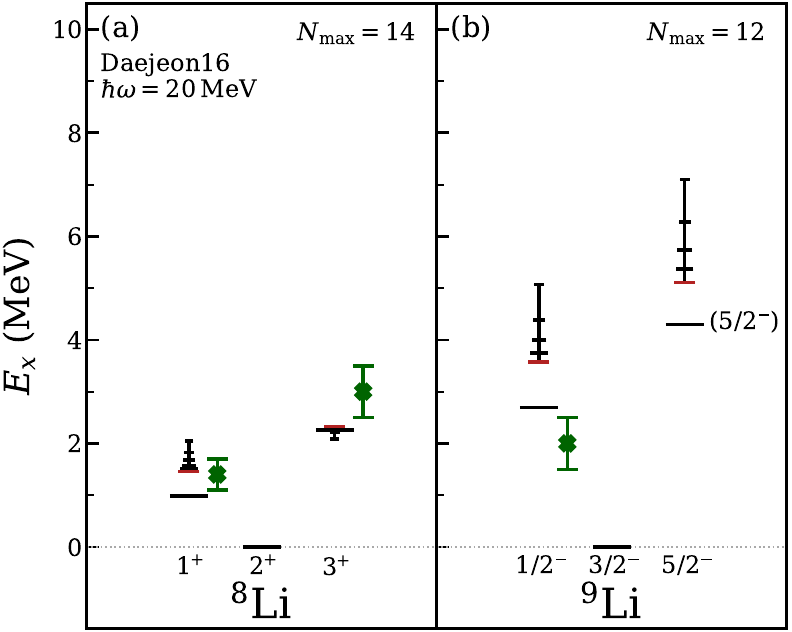}
\caption{Calculated excitation energies of low-lying states in
  (a)~$\isotope[8]{Li}$ and (b)~$\isotope[9]{Li}$, with angular momentum and
  parity as indicated at bottom, as obtained with the Daejeon16 interaction.
  Calculated values are shown at fixed $\hw=20\,\MeV$ and varying $\Nmax$
  (increasing symbol size), from $\Nmax=4$ to the maximum value indicated (at
  top).  Experimental energies~\cite{npa2004:008-010} are shown (horizontal line
  and error band) where available, as are the GFMC AV18+IL7
  predictions~\cite{pastore2013:qmc-em-alt9} (crosses) (see Table~III of
  Ref.~\cite{carlson2015:qmc-nuclear}).  }
\label{fig:excitation-teardrop-prediction}
\end{figure}

Returning to the $2^+\rightarrow1^+$ transition in $\isotope[8]{Li}$, the NCCI
calculations for the relevant dimensionless ratio are shown in
Fig.~\ref{fig:be2-norm-qp-scan-8li}.  For context, calculated and experimental
excitation energies of low-lying levels in $\isotope[8]{Li}$ are shown in
Fig.~\ref{fig:excitation-teardrop-prediction}(a).  We again compare results obtained
for the Daejeon16 [Fig.~\ref{fig:be2-norm-qp-scan-8li}(a)], JISP16
[Fig.~\ref{fig:be2-norm-qp-scan-8li}(b)], and LENPIC
[Fig.~\ref{fig:be2-norm-qp-scan-8li}(c)] interactions.

Focusing first on the Daejeon16 results
[Fig.~\ref{fig:be2-norm-qp-scan-8li}(a)], we see that taking the dimensionless
ratio $B(E2; 2^+\rightarrow 1^+)/[eQ(2^+)]^2$ rapidly eliminates the $\hw$ and
$\Nmax$ dependence, at the scale shown, even for modest $\Nmax$.  Calibrating to
the known $Q(2^+)=+3.14(2)\,\fm^2$~\cite{stone2016:e2-moments} yields the scale
at far right.  A ratio of $\approx0.18$, taken in conjunction with this
quadrupole moment, yields an estimated $B(E2; 2^+\rightarrow
1^+)\approx1.8\,e^2\fm^4$.

For the JISP16 interaction [Fig.~\ref{fig:be2-norm-qp-scan-8li}(b)], the
dimensionless
ratio exhibits greater $\hw$ dependence
than found for Daejeon16 [Fig.~\ref{fig:be2-norm-qp-scan-8li}(a)], especially
for lower $\Nmax$. Nonetheless, it appears to robustly converge
towards a result, $B(E2; 2^+\rightarrow 1^+)/[eQ(2^+)]^2\approx0.10$, in this
case lower by nearly a factor of two than obtained for Daejeon16.

For the LENPIC interaction [Fig.~\ref{fig:be2-norm-qp-scan-8li}(c)], taking the
dimensionless ratio tames the $\hw$ dependence, indeed, more effectively than
for JISP16 [Fig.~\ref{fig:be2-norm-qp-scan-8li}(b)].  There is still a slow but
steady increase with $\Nmax$ over much of the $\hw$ range.  Nonetheless, with
this caveat, the calculated ratio is again in the vicinity of
$0.10$.\footnote{The earlier NCCI calculations of Maris \textit{et
    al.}~\cite{maris2013:ncsm-chiral-a7-a8}, based on the chiral \nthreelo{}
  two-nucleon interaction of Entem and
  Machleidt~\cite{entem2003:chiral-nn-potl}, together with the \ntwolo{}
  three-nucleon interaction of Navr\'atil~\cite{navratil2007:local-chiral-3n},
  carried out using a basis with $\Nmax=8$ and $\hw=13\,\MeV$, and calculated
  with an
  Okubo-Lee-Suzuki~\cite{okubo1954:diagonalization-tamm-dancoff,suzuki1980:effective-interaction}
  renormalized effective interaction, give $Q(2^+)=2.648\,\fm^2$ and $B(E2;
  2^+\rightarrow 1^+)=0.714\,e^2\fm^4$, similarly yielding a ratio of $B(E2;
  2^+\rightarrow 1^+)/[eQ(2^+)]^2\approx0.10$.}

\begin{figure*}
\centering
\includegraphics[width=\ifproofpre{0.75}{0.95}\hsize]{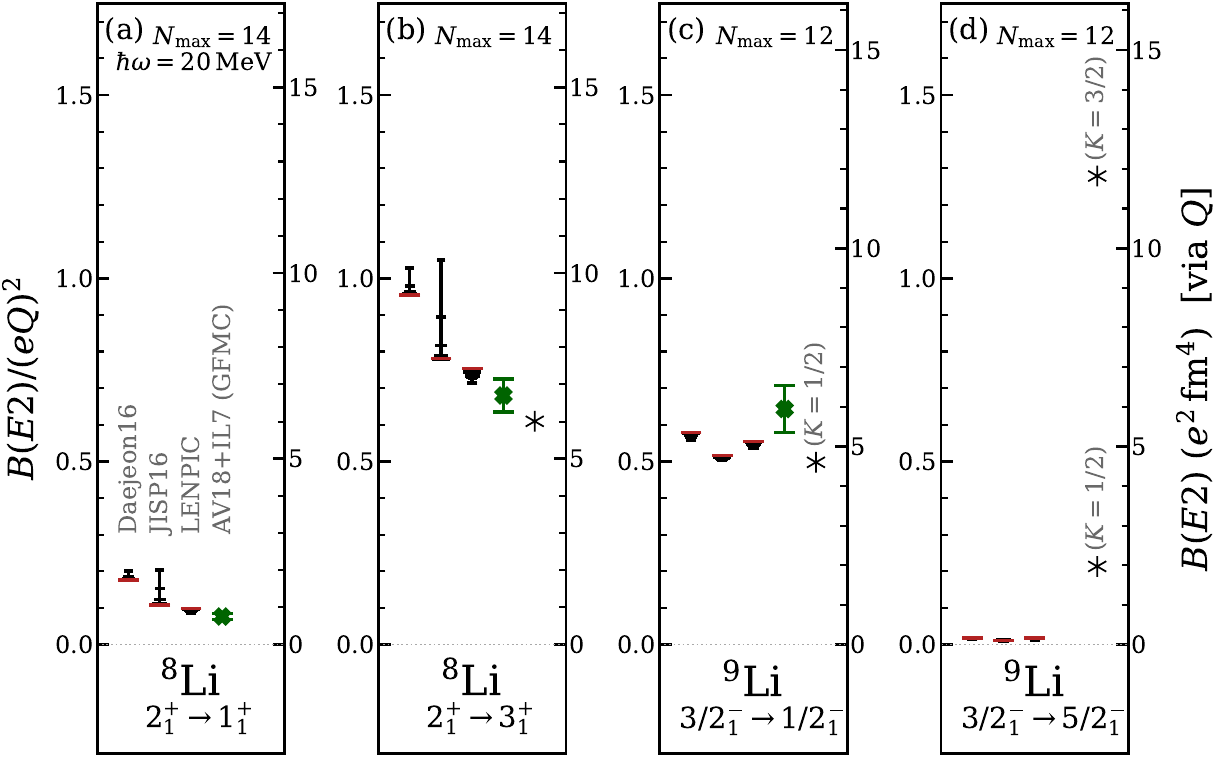}
\caption{Calculated ratios of the form $B(E2)/(eQ)^2$, for excitation to
  low-lying states in $\isotope[8]{Li}$ and $\isotope[9]{Li}$, obtained with the
  Daejeon16, JISP16, and LENPIC interactions (from left to right, for each
  transition).  Calculated values are shown at fixed $\hw=20\,\MeV$ and varying
  $\Nmax$ (increasing symbol size), from $\Nmax=4$ to the maximum value
  indicated (at top).  When calibrated to the experimental quadrupole
  moment~\cite{stone2016:e2-moments}, this ratio provides an estimate for the
  absolute $B(E2)$ (scale at right).  The experimental
  result~\cite{npa2004:008-010,stone2016:e2-moments} for the $\isotope[8]{Li}$
  $2^+ \rightarrow 1^+$ transition strength lies off scale. The GFMC AV18+IL7
  predictions~\cite{pastore2013:qmc-em-alt9} (crosses) and rotational ratios
  (asterisks) are shown.  }
\label{fig:be2-teardrop-prediction}
\end{figure*}

Thus, as summarized in Fig.~\ref{fig:be2-teardrop-prediction}(a), the NCCI
predictions show the ratio $B(E2)/(eQ)^2$ to depend upon the choice of
interaction, varying within the range $\approx 0.1\text{--}0.2$.  By way of
comparison, the GFMC calculation~\cite{pastore2013:qmc-em-alt9} gives $B(E2;
2^+\rightarrow 1^+)=0.83(7)\,e^2\fm^4$ and $Q(2^+)=+3.3(1)\,\fm^2$, which, recast
as a ratio, yield $B(E2; 2^+\rightarrow 1^+)/[eQ(2^+)]^2=0.076(8)$, similar in
scale to and marginally below these NCCI estimates.

That the \textit{ab initio} predictions for the $2^+\rightarrow1^+$ transition,
and in particular for the ratio to the squared quadrupole moment, show a greater
dependence upon the internucleon interaction than found above
(Sec.~\ref{sec:results-benchmark}) for the in-band rotational transitions in
$\isotope[7]{Li}$ and $\isotope[9]{Be}$ is perhaps not surprising.  One may take
the perspective that the $E2$ ratio is not ``constrained'' by the symmetry
considerations which apply to in-band transitions in an axially symmetric rotor
or, perhaps, an Elliott $\grpsu{3}$
rotor~\cite{elliott1958:su3-part1,*elliott1958:su3-part2,*elliott1963:su3-part3,*elliott1968:su3-part4,harvey1968:su3-shell,millener2001:light-nuclei,*millener2007:p-shell-hypernuclei}.
If the $2^+\rightarrow1^+$ transition is taken to be an interband transition,
rather, it is sensitive to the detailed microscopic structure of rotational
intrinsic states.  More generally, the transition involved is (predicted to be)
a weak (``noncollective'') transition, which might be expected to be sensitive,
\textit{e.g.}, in a shell model picture, to admixtures of different $p$-shell
configurations favored by the different interactions.

However, taken in conjunction with the known
$Q(2^+)=+3.14(2)\,\fm^2$~\cite{stone2016:e2-moments}, these \textit{ab initio}
results are all consistent with a modest strength of
$\approx1\text{--}2\,e^2\fm^4$ for the $2^+\rightarrow1^+$ transition, more than
an order of magnitude smaller than the experimental value of
$55(15)\,e^2\fm^4$~\cite{brown1991:8li-coulex,npa2004:008-010}.  It is thus of
particular interest to obtain confirmation of this reported strength.

It is interesting to contrast the results for this $2^+\rightarrow 1^+$
transition in $\isotope[8]{Li}$ with the results for the ostensibly in-band
$2^+\rightarrow3^+$ transition, shown in
Fig.~\ref{fig:be2-teardrop-prediction}(b).  In a rotational description, the
$3^+$ second excited state (a narrow resonance at $2.2\,\MeV$, just above the
neutron separation threshold) is naturally taken as a member of the $K=2$ ground
state band.  Experimentally, only the $M1$ partial decay width is
known~\cite{npa2004:008-010}, from a $\isotope[7]{Li}(n,\gamma)$
measurement~\cite{heil1998:7li-ngamma}, while a Coulomb excitation measurement for
the $E2$ strength would require neutron detection.  The
NCCI calculations, as obtained with the three different interactions, suggest
ratios $B(E2; 2^+\rightarrow 3^+)/[eQ(2^+)]^2$ in the range $\approx0.7$--$1.0$,
with the GFMC prediction~\cite{pastore2013:qmc-em-alt9} coming in at the low end
of this range, and the rotational ratio of $\approx0.609$ coming lower still.
In conjunction with the known quadrupole moment, the NCCI calculated ratios
yield a comparatively collective $B(E2;2^+\rightarrow 3^+)$ of
$\approx7$--$10\,e^2\fm^4$.

We conclude with NCCI predictions for the unmeasured $E2$ strengths from the
$3/2^-$ ground state of $\isotope[9]{Li}$ to the first two excited
states~\cite{npa2004:008-010}.  The only excited state below the neutron
threshold is a $1/2^-$ state at $\approx2.7\,\MeV$, while a resonance at
$\approx 4.3\,\MeV$, just above the neutron threshold, has tentative $(5/2^-)$
assignment.  This low-lying spectrum is consistent with the level ordering
obtained in the present NCCI calculations.  Calculated and experimental
excitation energies are shown in
Fig.~\ref{fig:excitation-teardrop-prediction}(b).

The NCCI predictions for the dimensionless ratio $B(E2;3/2^-\rightarrow
1/2^-)/[eQ(3/2^-)]^2$, shown in Fig.~\ref{fig:be2-teardrop-prediction}(c), are
robustly converged with respect to basis truncation.  The ratio is found to
depend modestly upon interaction, within the range $\approx0.5\text{--}0.6$.
Calibrating to the known ground-state quadrupole
moment~\cite{stone2016:e2-moments} yields strengths, depending upon interaction,
in the range $B(E2;3/2^-\rightarrow 1/2^-)\approx4.6\text{--}5.5\,e^2\fm^4$.
The GFMC AV18+IL7 predictions~\cite{pastore2013:qmc-em-alt9}, recast as a ratio,
give $0.64(6)$, which is roughly consistent with the ratios found in the NCCI
calculations.  However, on an absolute scale, the GFMC calculated
$Q(3/2^-)=-2.3(1)\,\fm^2$ underpredicts the experimental quadrupole moment by
$\approx24\%$, and the calculated $B(E2; 3/2^-\rightarrow
1/2^-)=3.40(17)\,e^2\fm^4$ is thus correspondingly lower than the above
estimates.

In a rotational description, it is not \textit{a priori} obvious whether this
transition should be interpreted as an in-band transition within a
Coriolis-staggered $K=1/2$ band, as in $\isotope[7]{Li}$
(Sec.~\ref{sec:results-benchmark}), or an interband transition between $K=3/2$
ground state and $K=1/2$ excited band heads.  The former interpretation would give an expected
rotational ratio of $B(E2;3/2_{K=1/2}\rightarrow
1/2_{K=1/2})/[eQ(3/2_{K=1/2})]^2\approx0.497$, as above for $\isotope[7]{Li}$,
while in the latter case the rotational prediction would depend on the ratio of
interband and in-band intrinsic matrix elements.  The \textit{ab initio} results
are roughly consistent with the $K=1/2$ in-band interpretation.

For the transition to the $5/2^-$ state, the NCCI calculations, shown in
Fig.~\ref{fig:be2-teardrop-prediction}(d), give $B(E2;3/2^-\rightarrow
5/2^-)/[eQ(3/2^-)]^2\approx0.01\text{--}0.02$, depending upon choice of
interaction, yielding a comparatively weak $B(E2;3/2^-\rightarrow
5/2^-)\approx0.1\text{--}0.2\,e^2\fm^4$.  The \textit{ab initio} predicted ratio
is not conducive to an interpretation of this transition as a rotational in-band
transition, whether within a Coriolis-staggered $K=1/2$ band, for which
$B(E2;3/2_{K=1/2}\rightarrow 5/2_{K=1/2})/[eQ(3/2_{K=1/2})]^2\approx0.213$, or
within a $K=3/2$ band built on the ground state, for which
$B(E2;3/2_{K=3/2}\rightarrow 5/2_{K=3/2})/[eQ(3/2_{K=3/2})]^2\approx1.279$.
 \section{Conclusion}

Although meaningful, converged predictions for $E2$ observables are elusive in
\textit{ab initio} NCCI calculations, calculated $E2$ observables are
correlated, presumably due to their common dependence on the truncation of the
long-distance tails of the wave functions.  For the ground-state quadrupole
moment and low-lying transitions, we demonstrate that much of this systematic
truncation error cancels out in dimensionless ratios of the form $B(E2)/(eQ)^2$,
allowing robust predictions to be obtained.  Calibrating to the known
ground-state quadrupole moment then provides an $E2$ strength estimate on an
absolute scale.

For the rotational in-band transitions in $\isotope[7]{Li}$ and
$\isotope[9]{Be}$, there is general agreement, in the $B(E2)/(eQ)^2$ ratios,
between the predictions obtained across several choices for the internucleon
interaction.  These calculated values, like the experimental ratios and GFMC
predictions, are approximately consistent with the simple axial rotor model, and
calibrating to the ground-state quadrupole moment reproduces the experimentally
observed $E2$ enhancement.  For the $2^+\rightarrow1^+$ transition in
$\isotope[8]{Li}$, which is not naturally interpreted as a rotational in-band
transition, robust $\textit{ab initio}$ predictions are made for the ratio
$B(E2)/(eQ)^2$, showing modest dependence on the choice of internucleon
interaction, and reinforcing the severe tension between \textit{ab initio}
theory~\cite{pastore2013:qmc-em-alt9} and
experiment~\cite{brown1991:8li-coulex,npa2004:008-010} for this transition.
Finally, we provide robust \textit{ab initio} predictions for the ratio
$B(E2)/(eQ)^2$, and thus, by normalization to the experimental ground state
quadrupole moment, estimates for unmeasured $E2$ strengths to the low-lying
$3^+$ resonance of $\isotope[8]{Li}$ and to low-lying states of
$\isotope[9]{Li}$.

\begin{acknowledgments}
  We thank Tan Ahn, Samuel L.~Henderson, Pieter Maris, and Anna E.~McCoy for valuable
  discussions,
  James P.~Vary, Ik Jae Shin, and Youngman Kim for sharing illuminating
  results on ratios of observables,
  and Colin V.~Coane, Jakub Herko, and Zhou Zhou for comments
  on the manuscript.  This material is based upon work supported by the
  U.S.~Department of Energy, Office of Science, under Award
  No.~DE-FG02-95ER40934.  This research used resources of the National Energy
  Research Scientific Computing Center (NERSC), a DOE Office of Science User
  Facility supported by the Office of Science of the U.S.~Department of Energy
  under Contract No.~DE-AC02-05CH11231, using NERSC award NP-ERCAP0020422.
\end{acknowledgments}

\bibliographystyle{apsrev4-2}

\begin{thebibliography}{58}\makeatletter
\providecommand \@ifxundefined [1]{\@ifx{#1\undefined}
}\providecommand \@ifnum [1]{\ifnum #1\expandafter \@firstoftwo
 \else \expandafter \@secondoftwo
 \fi
}\providecommand \@ifx [1]{\ifx #1\expandafter \@firstoftwo
 \else \expandafter \@secondoftwo
 \fi
}\providecommand \natexlab [1]{#1}\providecommand \enquote  [1]{``#1''}\providecommand \bibnamefont  [1]{#1}\providecommand \bibfnamefont [1]{#1}\providecommand \citenamefont [1]{#1}\providecommand \href@noop [0]{\@secondoftwo}\providecommand \href [0]{\begingroup \@sanitize@url \@href}\providecommand \@href[1]{\@@startlink{#1}\@@href}\providecommand \@@href[1]{\endgroup#1\@@endlink}\providecommand \@sanitize@url [0]{\catcode `\\12\catcode `\$12\catcode
  `\&12\catcode `\#12\catcode `\^12\catcode `\_12\catcode `\%12\relax}\providecommand \@@startlink[1]{}\providecommand \@@endlink[0]{}\providecommand \url  [0]{\begingroup\@sanitize@url \@url }\providecommand \@url [1]{\endgroup\@href {#1}{\urlprefix }}\providecommand \urlprefix  [0]{URL }\providecommand \Eprint [0]{\href }\providecommand \doibase [0]{https://doi.org/}\providecommand \selectlanguage [0]{\@gobble}\providecommand \bibinfo  [0]{\@secondoftwo}\providecommand \bibfield  [0]{\@secondoftwo}\providecommand \translation [1]{[#1]}\providecommand \BibitemOpen [0]{}\providecommand \bibitemStop [0]{}\providecommand \bibitemNoStop [0]{.\EOS\space}\providecommand \EOS [0]{\spacefactor3000\relax}\providecommand \BibitemShut  [1]{\csname bibitem#1\endcsname}\let\auto@bib@innerbib\@empty
\bibitem [{\citenamefont {Bohr}\ and\ \citenamefont
  {Mottelson}(1998)}]{bohr1998:v2}\BibitemOpen
  \bibfield  {author} {\bibinfo {author} {\bibfnamefont {A.}~\bibnamefont
  {Bohr}}\ and\ \bibinfo {author} {\bibfnamefont {B.~R.}\ \bibnamefont
  {Mottelson}},\ }\href@noop {} {\emph {\bibinfo {title} {Nuclear
  Structure}}},\ Vol.~\bibinfo {volume} {2}\ (\bibinfo  {publisher} {World
  Scientific},\ \bibinfo {address} {Singapore},\ \bibinfo {year}
  {1998})\BibitemShut {NoStop}\bibitem [{\citenamefont {Casten}(2000)}]{casten2000:ns}\BibitemOpen
  \bibfield  {author} {\bibinfo {author} {\bibfnamefont {R.~F.}\ \bibnamefont
  {Casten}},\ }\href@noop {} {\emph {\bibinfo {title} {Nuclear Structure from a
  Simple Perspective}}},\ \bibinfo {edition} {2nd}\ ed.,\ \bibinfo {series}
  {Oxford Studies in Nuclear Physics}\ No.~\bibinfo {number} {23}\ (\bibinfo
  {publisher} {Oxford University Press},\ \bibinfo {address} {Oxford},\
  \bibinfo {year} {2000})\BibitemShut {NoStop}\bibitem [{\citenamefont {Rowe}(2010)}]{rowe2010:collective-motion}\BibitemOpen
  \bibfield  {author} {\bibinfo {author} {\bibfnamefont {D.~J.}\ \bibnamefont
  {Rowe}},\ }\href@noop {} {\emph {\bibinfo {title} {Nuclear Collective Motion:
  Models and Theory}}}\ (\bibinfo  {publisher} {World Scientific},\ \bibinfo
  {address} {Singapore},\ \bibinfo {year} {2010})\BibitemShut {NoStop}\bibitem [{\citenamefont {Bogner}\ \emph {et~al.}(2008)\citenamefont {Bogner},
  \citenamefont {Furnstahl}, \citenamefont {Maris}, \citenamefont {Perry},
  \citenamefont {Schwenk},\ and\ \citenamefont
  {Vary}}]{bogner2008:ncsm-converg-2N}\BibitemOpen
  \bibfield  {author} {\bibinfo {author} {\bibfnamefont {S.~K.}\ \bibnamefont
  {Bogner}}, \bibinfo {author} {\bibfnamefont {R.~J.}\ \bibnamefont
  {Furnstahl}}, \bibinfo {author} {\bibfnamefont {P.}~\bibnamefont {Maris}},
  \bibinfo {author} {\bibfnamefont {R.~J.}\ \bibnamefont {Perry}}, \bibinfo
  {author} {\bibfnamefont {A.}~\bibnamefont {Schwenk}},\ and\ \bibinfo {author}
  {\bibfnamefont {J.}~\bibnamefont {Vary}},\ }\bibfield  {title} {\bibinfo
  {title} {Convergence in the no-core shell model with low-momentum two-nucleon
  interactions},\ }\href {https://doi.org/10.1016/j.nuclphysa.2007.12.008}
  {\bibfield  {journal} {\bibinfo  {journal} {Nucl. Phys. A}\ }\textbf
  {\bibinfo {volume} {801}},\ \bibinfo {pages} {21} (\bibinfo {year}
  {2008})}\BibitemShut {NoStop}\bibitem [{\citenamefont {Maris}\ and\ \citenamefont
  {Vary}(2013)}]{maris2013:ncsm-pshell}\BibitemOpen
  \bibfield  {author} {\bibinfo {author} {\bibfnamefont {P.}~\bibnamefont
  {Maris}}\ and\ \bibinfo {author} {\bibfnamefont {J.~P.}\ \bibnamefont
  {Vary}},\ }\bibfield  {title} {\bibinfo {title} {\textit{Ab initio} nuclear
  structure calculations of $p$-shell nuclei with {JISP16}},\ }\href
  {https://doi.org/10.1142/S0218301313300166} {\bibfield  {journal} {\bibinfo
  {journal} {Int. J. Mod. Phys. E}\ }\textbf {\bibinfo {volume} {22}},\
  \bibinfo {pages} {1330016} (\bibinfo {year} {2013})}\BibitemShut {NoStop}\bibitem [{\citenamefont {Odell}\ \emph {et~al.}(2016)\citenamefont {Odell},
  \citenamefont {Papenbrock},\ and\ \citenamefont
  {Platter}}]{odell2016:ir-extrap-quadrupole}\BibitemOpen
  \bibfield  {author} {\bibinfo {author} {\bibfnamefont {D.}~\bibnamefont
  {Odell}}, \bibinfo {author} {\bibfnamefont {T.}~\bibnamefont {Papenbrock}},\
  and\ \bibinfo {author} {\bibfnamefont {L.}~\bibnamefont {Platter}},\
  }\bibfield  {title} {\bibinfo {title} {Infrared extrapolations of quadrupole
  moments and transitions},\ }\href
  {https://doi.org/10.1103/PhysRevC.93.044331} {\bibfield  {journal} {\bibinfo
  {journal} {Phys. Rev. C}\ }\textbf {\bibinfo {volume} {93}},\ \bibinfo
  {pages} {044331} (\bibinfo {year} {2016})}\BibitemShut {NoStop}\bibitem [{\citenamefont {Barrett}\ \emph {et~al.}(2013)\citenamefont
  {Barrett}, \citenamefont {Navr\'{a}til},\ and\ \citenamefont
  {Vary}}]{barrett2013:ncsm}\BibitemOpen
  \bibfield  {author} {\bibinfo {author} {\bibfnamefont {B.~R.}\ \bibnamefont
  {Barrett}}, \bibinfo {author} {\bibfnamefont {P.}~\bibnamefont
  {Navr\'{a}til}},\ and\ \bibinfo {author} {\bibfnamefont {J.~P.}\ \bibnamefont
  {Vary}},\ }\bibfield  {title} {\bibinfo {title} {\textit{Ab initio} no core
  shell model},\ }\href {https://doi.org/10.1016/j.ppnp.2012.10.003} {\bibfield
   {journal} {\bibinfo  {journal} {Prog. Part. Nucl. Phys.}\ }\textbf {\bibinfo
  {volume} {69}},\ \bibinfo {pages} {131} (\bibinfo {year} {2013})}\BibitemShut
  {NoStop}\bibitem [{\citenamefont {Roth}\ and\ \citenamefont
  {Navr\'atil}(2007)}]{roth2007:it-ncsm-40ca}\BibitemOpen
  \bibfield  {author} {\bibinfo {author} {\bibfnamefont {R.}~\bibnamefont
  {Roth}}\ and\ \bibinfo {author} {\bibfnamefont {P.}~\bibnamefont
  {Navr\'atil}},\ }\bibfield  {title} {\bibinfo {title} {\textit{Ab initio}
  study of $\isotope[40]{Ca}$ with an importance-truncated no-core shell
  model},\ }\href {https://doi.org/10.1103/PhysRevLett.99.092501} {\bibfield
  {journal} {\bibinfo  {journal} {Phys. Rev. Lett.}\ }\textbf {\bibinfo
  {volume} {99}},\ \bibinfo {pages} {092501} (\bibinfo {year}
  {2007})}\BibitemShut {NoStop}\bibitem [{\citenamefont {Dytrych}\ \emph {et~al.}(2008)\citenamefont
  {Dytrych}, \citenamefont {Sviratcheva}, \citenamefont {Draayer},
  \citenamefont {Bahri},\ and\ \citenamefont {Vary}}]{dytrych2008:sp-ncsm}\BibitemOpen
  \bibfield  {author} {\bibinfo {author} {\bibfnamefont {T.}~\bibnamefont
  {Dytrych}}, \bibinfo {author} {\bibfnamefont {K.~D.}\ \bibnamefont
  {Sviratcheva}}, \bibinfo {author} {\bibfnamefont {J.~P.}\ \bibnamefont
  {Draayer}}, \bibinfo {author} {\bibfnamefont {C.}~\bibnamefont {Bahri}},\
  and\ \bibinfo {author} {\bibfnamefont {J.~P.}\ \bibnamefont {Vary}},\
  }\bibfield  {title} {\bibinfo {title} {\textit{Ab initio} symplectic no-core
  shell model},\ }\href {https://doi.org/10.1088/0954-3899/35/12/123101}
  {\bibfield  {journal} {\bibinfo  {journal} {J. Phys. G}\ }\textbf {\bibinfo
  {volume} {35}},\ \bibinfo {pages} {123101} (\bibinfo {year}
  {2008})}\BibitemShut {NoStop}\bibitem [{\citenamefont {Dytrych}\ \emph {et~al.}(2016)\citenamefont
  {Dytrych}, \citenamefont {Maris}, \citenamefont {Launey}, \citenamefont
  {Draayer}, \citenamefont {Vary}, \citenamefont {Langr}, \citenamefont
  {Saule}, \citenamefont {Caprio}, \citenamefont {Catalyurek},\ and\
  \citenamefont {Sosonkina}}]{dytrych2016:su3ncsm-12c-efficacy}\BibitemOpen
  \bibfield  {author} {\bibinfo {author} {\bibfnamefont {T.}~\bibnamefont
  {Dytrych}}, \bibinfo {author} {\bibfnamefont {P.}~\bibnamefont {Maris}},
  \bibinfo {author} {\bibfnamefont {K.~D.}\ \bibnamefont {Launey}}, \bibinfo
  {author} {\bibfnamefont {J.~P.}\ \bibnamefont {Draayer}}, \bibinfo {author}
  {\bibfnamefont {J.~P.}\ \bibnamefont {Vary}}, \bibinfo {author}
  {\bibfnamefont {D.}~\bibnamefont {Langr}}, \bibinfo {author} {\bibfnamefont
  {E.}~\bibnamefont {Saule}}, \bibinfo {author} {\bibfnamefont {M.~A.}\
  \bibnamefont {Caprio}}, \bibinfo {author} {\bibfnamefont {U.}~\bibnamefont
  {Catalyurek}},\ and\ \bibinfo {author} {\bibfnamefont {M.}~\bibnamefont
  {Sosonkina}},\ }\bibfield  {title} {\bibinfo {title} {Efficacy of the
  $\grpsu{3}$ scheme for \textit{ab initio} large-scale calculations beyond the
  lightest nuclei},\ }\href {https://doi.org/10.1016/j.cpc.2016.06.006}
  {\bibfield  {journal} {\bibinfo  {journal} {Comput. Phys. Commun.}\ }\textbf
  {\bibinfo {volume} {207}},\ \bibinfo {pages} {202} (\bibinfo {year}
  {2016})}\BibitemShut {NoStop}\bibitem [{\citenamefont {Vorabbi}\ \emph {et~al.}(2019)\citenamefont
  {Vorabbi}, \citenamefont {Navr\'atil}, \citenamefont {Quaglioni},\ and\
  \citenamefont {Hupin}}]{vorabbi2019:7be-7li-ncsmc}\BibitemOpen
  \bibfield  {author} {\bibinfo {author} {\bibfnamefont {M.}~\bibnamefont
  {Vorabbi}}, \bibinfo {author} {\bibfnamefont {P.}~\bibnamefont {Navr\'atil}},
  \bibinfo {author} {\bibfnamefont {S.}~\bibnamefont {Quaglioni}},\ and\
  \bibinfo {author} {\bibfnamefont {G.}~\bibnamefont {Hupin}},\ }\bibfield
  {title} {\bibinfo {title} {$\isotope[7]{Be}$ and $\isotope[7]{Li}$ nuclei
  within the no-core shell model with continuum},\ }\href
  {https://doi.org/10.1103/PhysRevC.100.024304} {\bibfield  {journal} {\bibinfo
   {journal} {Phys. Rev. C}\ }\textbf {\bibinfo {volume} {100}},\ \bibinfo
  {pages} {024304} (\bibinfo {year} {2019})}\BibitemShut {NoStop}\bibitem [{\citenamefont {McCoy}\ \emph {et~al.}(2020)\citenamefont {McCoy},
  \citenamefont {Caprio}, \citenamefont {Dytrych},\ and\ \citenamefont
  {Fasano}}]{mccoy2020:spfamilies}\BibitemOpen
  \bibfield  {author} {\bibinfo {author} {\bibfnamefont {A.~E.}\ \bibnamefont
  {McCoy}}, \bibinfo {author} {\bibfnamefont {M.~A.}\ \bibnamefont {Caprio}},
  \bibinfo {author} {\bibfnamefont {T.}~\bibnamefont {Dytrych}},\ and\ \bibinfo
  {author} {\bibfnamefont {P.~J.}\ \bibnamefont {Fasano}},\ }\bibfield  {title}
  {\bibinfo {title} {Emergent {$\grpsptr$} dynamical symmetry in the nuclear
  many-body system from an \emph{ab initio} description},\ }\href
  {https://doi.org/10.1103/PhysRevLett.125.102505} {\bibfield  {journal}
  {\bibinfo  {journal} {Phys. Rev. Lett.}\ }\textbf {\bibinfo {volume} {125}},\
  \bibinfo {pages} {102505} (\bibinfo {year} {2020})}\BibitemShut {NoStop}\bibitem [{\citenamefont {Fasano}\ \emph {et~al.}(2022)\citenamefont {Fasano},
  \citenamefont {Constantinou}, \citenamefont {Caprio}, \citenamefont {Maris},\
  and\ \citenamefont {Vary}}]{fasano2022:natorb}\BibitemOpen
  \bibfield  {author} {\bibinfo {author} {\bibfnamefont {P.~J.}\ \bibnamefont
  {Fasano}}, \bibinfo {author} {\bibfnamefont {{\mbox{Ch}}.}~\bibnamefont
  {Constantinou}}, \bibinfo {author} {\bibfnamefont {M.~A.}\ \bibnamefont
  {Caprio}}, \bibinfo {author} {\bibfnamefont {P.}~\bibnamefont {Maris}},\ and\
  \bibinfo {author} {\bibfnamefont {J.~P.}\ \bibnamefont {Vary}},\ }\bibfield
  {title} {\bibinfo {title} {Natural orbitals for the \textit{ab initio}
  no-core configuration interaction approach},\ }\href
  {https://doi.org/10.1103/PhysRevC.105.054301} {\bibfield  {journal} {\bibinfo
   {journal} {Phys. Rev. C}\ }\textbf {\bibinfo {volume} {105}},\ \bibinfo
  {pages} {054301} (\bibinfo {year} {2022})}\BibitemShut {NoStop}\bibitem [{\citenamefont {Caprio}\ \emph {et~al.}(2013)\citenamefont {Caprio},
  \citenamefont {Maris},\ and\ \citenamefont {Vary}}]{caprio2013:berotor}\BibitemOpen
  \bibfield  {author} {\bibinfo {author} {\bibfnamefont {M.~A.}\ \bibnamefont
  {Caprio}}, \bibinfo {author} {\bibfnamefont {P.}~\bibnamefont {Maris}},\ and\
  \bibinfo {author} {\bibfnamefont {J.~P.}\ \bibnamefont {Vary}},\ }\bibfield
  {title} {\bibinfo {title} {Emergence of rotational bands in \textit{ab
  initio} no-core configuration interaction calculations of light nuclei},\
  }\href {https://doi.org/10.1016/j.physletb.2012.12.064} {\bibfield  {journal}
  {\bibinfo  {journal} {Phys. Lett. B}\ }\textbf {\bibinfo {volume} {719}},\
  \bibinfo {pages} {179} (\bibinfo {year} {2013})}\BibitemShut {NoStop}\bibitem [{\citenamefont {Maris}\ \emph {et~al.}(2015)\citenamefont {Maris},
  \citenamefont {Caprio},\ and\ \citenamefont {Vary}}]{maris2015:berotor2}\BibitemOpen
  \bibfield  {author} {\bibinfo {author} {\bibfnamefont {P.}~\bibnamefont
  {Maris}}, \bibinfo {author} {\bibfnamefont {M.~A.}\ \bibnamefont {Caprio}},\
  and\ \bibinfo {author} {\bibfnamefont {J.~P.}\ \bibnamefont {Vary}},\
  }\bibfield  {title} {\bibinfo {title} {Emergence of rotational bands in
  \textit{ab initio} no-core configuration interaction calculations of the
  $\isotope{Be}$ isotopes},\ }\href
  {https://doi.org/10.1103/PhysRevC.91.014310} {\bibfield  {journal} {\bibinfo
  {journal} {Phys. Rev. C}\ }\textbf {\bibinfo {volume} {91}},\ \bibinfo
  {pages} {014310} (\bibinfo {year} {2015})}\BibitemShut {NoStop}\bibitem [{\citenamefont {Maris}\ \emph {et~al.}(2019)\citenamefont {Maris},
  \citenamefont {Caprio},\ and\ \citenamefont
  {Vary}}]{maris2019:berotor2-ERRATUM}\BibitemOpen
  \bibfield  {author} {\bibinfo {author} {\bibfnamefont {P.}~\bibnamefont
  {Maris}}, \bibinfo {author} {\bibfnamefont {M.~A.}\ \bibnamefont {Caprio}},\
  and\ \bibinfo {author} {\bibfnamefont {J.~P.}\ \bibnamefont {Vary}},\
  }\bibfield  {title} {\bibinfo {title} {Erratum: Emergence of rotational bands
  in \textit{ab initio} no-core configuration interaction calculations of the
  $\isotope{Be}$ isotopes},\ }\href
  {https://doi.org/10.1103/PhysRevC.99.029902} {\bibfield  {journal} {\bibinfo
  {journal} {Phys. Rev. C}\ }\textbf {\bibinfo {volume} {99}},\ \bibinfo
  {pages} {029902(E)} (\bibinfo {year} {2019})}\BibitemShut {NoStop}\bibitem [{\citenamefont {Calci}\ and\ \citenamefont
  {Roth}(2016)}]{calci2016:observable-correlations-chiral}\BibitemOpen
  \bibfield  {author} {\bibinfo {author} {\bibfnamefont {A.}~\bibnamefont
  {Calci}}\ and\ \bibinfo {author} {\bibfnamefont {R.}~\bibnamefont {Roth}},\
  }\bibfield  {title} {\bibinfo {title} {Sensitivities and correlations of
  nuclear structure observables emerging from chiral interactions},\ }\href
  {https://doi.org/10.1103/PhysRevC.94.014322} {\bibfield  {journal} {\bibinfo
  {journal} {Phys. Rev. C}\ }\textbf {\bibinfo {volume} {94}},\ \bibinfo
  {pages} {014322} (\bibinfo {year} {2016})}\BibitemShut {NoStop}\bibitem [{\citenamefont {Henderson}\ \emph {et~al.}(2019)\citenamefont
  {Henderson}, \citenamefont {Ahn}, \citenamefont {Caprio}, \citenamefont
  {Fasano}, \citenamefont {Simon}, \citenamefont {Tan}, \citenamefont
  {O'Malley}, \citenamefont {Allen}, \citenamefont {Bardayan}, \citenamefont
  {Blankstein}, \citenamefont {Frentz}, \citenamefont {Hall}, \citenamefont
  {Kolata}, \citenamefont {McCoy}, \citenamefont {Moylan}, \citenamefont
  {Reingold}, \citenamefont {Strauss},\ and\ \citenamefont
  {Torres-Isea}}]{henderson2019:7be-coulex}\BibitemOpen
  \bibfield  {author} {\bibinfo {author} {\bibfnamefont {S.~L.}\ \bibnamefont
  {Henderson}}, \bibinfo {author} {\bibfnamefont {T.}~\bibnamefont {Ahn}},
  \bibinfo {author} {\bibfnamefont {M.~A.}\ \bibnamefont {Caprio}}, \bibinfo
  {author} {\bibfnamefont {P.~J.}\ \bibnamefont {Fasano}}, \bibinfo {author}
  {\bibfnamefont {A.}~\bibnamefont {Simon}}, \bibinfo {author} {\bibfnamefont
  {W.}~\bibnamefont {Tan}}, \bibinfo {author} {\bibfnamefont {P.}~\bibnamefont
  {O'Malley}}, \bibinfo {author} {\bibfnamefont {J.}~\bibnamefont {Allen}},
  \bibinfo {author} {\bibfnamefont {D.~W.}\ \bibnamefont {Bardayan}}, \bibinfo
  {author} {\bibfnamefont {D.}~\bibnamefont {Blankstein}}, \bibinfo {author}
  {\bibfnamefont {B.}~\bibnamefont {Frentz}}, \bibinfo {author} {\bibfnamefont
  {M.~R.}\ \bibnamefont {Hall}}, \bibinfo {author} {\bibfnamefont {J.~J.}\
  \bibnamefont {Kolata}}, \bibinfo {author} {\bibfnamefont {A.~E.}\
  \bibnamefont {McCoy}}, \bibinfo {author} {\bibfnamefont {S.}~\bibnamefont
  {Moylan}}, \bibinfo {author} {\bibfnamefont {C.~S.}\ \bibnamefont
  {Reingold}}, \bibinfo {author} {\bibfnamefont {S.~Y.}\ \bibnamefont
  {Strauss}},\ and\ \bibinfo {author} {\bibfnamefont {R.~O.}\ \bibnamefont
  {Torres-Isea}},\ }\bibfield  {title} {\bibinfo {title} {First measurement of
  the {$B(E2; 3/2^- \rightarrow 1/2^-)$} transition strength in
  {$\isotope[7]{Be}$}: Testing \textit{ab initio} predictions for {$A=7$}
  nuclei},\ }\href {https://doi.org/10.1103/PhysRevC.99.064320} {\bibfield
  {journal} {\bibinfo  {journal} {Phys. Rev. C}\ }\textbf {\bibinfo {volume}
  {99}},\ \bibinfo {pages} {064320} (\bibinfo {year} {2019})}\BibitemShut
  {NoStop}\bibitem [{\citenamefont {Caprio}\ \emph {et~al.}(2019)\citenamefont {Caprio},
  \citenamefont {Fasano}, \citenamefont {McCoy}, \citenamefont {Maris},\ and\
  \citenamefont {Vary}}]{caprio2019:bebands-sdanca19}\BibitemOpen
  \bibfield  {author} {\bibinfo {author} {\bibfnamefont {M.~A.}\ \bibnamefont
  {Caprio}}, \bibinfo {author} {\bibfnamefont {P.~J.}\ \bibnamefont {Fasano}},
  \bibinfo {author} {\bibfnamefont {A.~E.}\ \bibnamefont {McCoy}}, \bibinfo
  {author} {\bibfnamefont {P.}~\bibnamefont {Maris}},\ and\ \bibinfo {author}
  {\bibfnamefont {J.~P.}\ \bibnamefont {Vary}},\ }\bibfield  {title} {\bibinfo
  {title} {\textit{Ab initio} rotation in {$\isotope[10]{Be}$}},\ }\href
  {https://www.bjp-bg.com/paper.php?id=1208} {\bibfield  {journal} {\bibinfo
  {journal} {Bulg. J. Phys.}\ }\textbf {\bibinfo {volume} {46}},\ \bibinfo
  {pages} {445} (\bibinfo {year} {2019})}\BibitemShut {NoStop}\bibitem [{\citenamefont {Caprio}\ \emph {et~al.}(2020)\citenamefont {Caprio},
  \citenamefont {Fasano}, \citenamefont {Maris}, \citenamefont {McCoy},\ and\
  \citenamefont {Vary}}]{caprio2020:bebands}\BibitemOpen
  \bibfield  {author} {\bibinfo {author} {\bibfnamefont {M.~A.}\ \bibnamefont
  {Caprio}}, \bibinfo {author} {\bibfnamefont {P.~J.}\ \bibnamefont {Fasano}},
  \bibinfo {author} {\bibfnamefont {P.}~\bibnamefont {Maris}}, \bibinfo
  {author} {\bibfnamefont {A.~E.}\ \bibnamefont {McCoy}},\ and\ \bibinfo
  {author} {\bibfnamefont {J.~P.}\ \bibnamefont {Vary}},\ }\bibfield  {title}
  {\bibinfo {title} {Probing \textit{ab initio} emergence of nuclear
  rotation},\ }\href {https://doi.org/10.1140/epja/s10050-020-00112-0}
  {\bibfield  {journal} {\bibinfo  {journal} {Eur. Phys. J. A}\ }\textbf
  {\bibinfo {volume} {56}},\ \bibinfo {pages} {120} (\bibinfo {year}
  {2020})}\BibitemShut {NoStop}\bibitem [{\citenamefont {Caprio}\ \emph {et~al.}(2021)\citenamefont {Caprio},
  \citenamefont {Fasano}, \citenamefont {Maris},\ and\ \citenamefont
  {McCoy}}]{caprio2021:emratio}\BibitemOpen
  \bibfield  {author} {\bibinfo {author} {\bibfnamefont {M.~A.}\ \bibnamefont
  {Caprio}}, \bibinfo {author} {\bibfnamefont {P.~J.}\ \bibnamefont {Fasano}},
  \bibinfo {author} {\bibfnamefont {P.}~\bibnamefont {Maris}},\ and\ \bibinfo
  {author} {\bibfnamefont {A.~E.}\ \bibnamefont {McCoy}},\ }\bibfield  {title}
  {\bibinfo {title} {Quadrupole moments and proton-neutron structure in
  $p$-shell mirror nuclei},\ }\href
  {https://doi.org/10.1103/PhysRevC.104.034319} {\bibfield  {journal} {\bibinfo
   {journal} {Phys. Rev. C}\ }\textbf {\bibinfo {volume} {104}},\ \bibinfo
  {pages} {034319} (\bibinfo {year} {2021})}\BibitemShut {NoStop}\bibitem [{\citenamefont {Stone}(2016)}]{stone2016:e2-moments}\BibitemOpen
  \bibfield  {author} {\bibinfo {author} {\bibfnamefont {N.~J.}\ \bibnamefont
  {Stone}},\ }\bibfield  {title} {\bibinfo {title} {Table of nuclear electric
  quadrupole moments},\ }\href {https://doi.org/10.1016/j.adt.2015.12.002}
  {\bibfield  {journal} {\bibinfo  {journal} {At. Data Nucl. Data Tables}\
  }\textbf {\bibinfo {volume} {111--112}},\ \bibinfo {pages} {1} (\bibinfo
  {year} {2016})}\BibitemShut {NoStop}\bibitem [{\citenamefont {Tilley}\ \emph {et~al.}(2002)\citenamefont {Tilley},
  \citenamefont {Cheves}, \citenamefont {Godwin}, \citenamefont {Hale},
  \citenamefont {Hofmann}, \citenamefont {Kelley}, \citenamefont {Sheu},\ and\
  \citenamefont {Weller}}]{npa2002:005-007}\BibitemOpen
  \bibfield  {author} {\bibinfo {author} {\bibfnamefont {D.~R.}\ \bibnamefont
  {Tilley}}, \bibinfo {author} {\bibfnamefont {C.~M.}\ \bibnamefont {Cheves}},
  \bibinfo {author} {\bibfnamefont {J.~L.}\ \bibnamefont {Godwin}}, \bibinfo
  {author} {\bibfnamefont {G.~M.}\ \bibnamefont {Hale}}, \bibinfo {author}
  {\bibfnamefont {H.~M.}\ \bibnamefont {Hofmann}}, \bibinfo {author}
  {\bibfnamefont {J.~H.}\ \bibnamefont {Kelley}}, \bibinfo {author}
  {\bibfnamefont {C.~G.}\ \bibnamefont {Sheu}},\ and\ \bibinfo {author}
  {\bibfnamefont {H.~R.}\ \bibnamefont {Weller}},\ }\bibfield  {title}
  {\bibinfo {title} {Energy levels of light nuclei {$A = 5, 6, 7$}},\ }\href
  {https://doi.org/10.1016/S0375-9474(02)00597-3} {\bibfield  {journal}
  {\bibinfo  {journal} {Nucl. Phys. A}\ }\textbf {\bibinfo {volume} {708}},\
  \bibinfo {pages} {3} (\bibinfo {year} {2002})}\BibitemShut {NoStop}\bibitem [{\citenamefont {Tilley}\ \emph {et~al.}(2004)\citenamefont {Tilley},
  \citenamefont {Kelley}, \citenamefont {Godwin}, \citenamefont {Millener},
  \citenamefont {Purcell}, \citenamefont {Sheu},\ and\ \citenamefont
  {Weller}}]{npa2004:008-010}\BibitemOpen
  \bibfield  {author} {\bibinfo {author} {\bibfnamefont {D.~R.}\ \bibnamefont
  {Tilley}}, \bibinfo {author} {\bibfnamefont {J.~H.}\ \bibnamefont {Kelley}},
  \bibinfo {author} {\bibfnamefont {J.~L.}\ \bibnamefont {Godwin}}, \bibinfo
  {author} {\bibfnamefont {D.~J.}\ \bibnamefont {Millener}}, \bibinfo {author}
  {\bibfnamefont {J.~E.}\ \bibnamefont {Purcell}}, \bibinfo {author}
  {\bibfnamefont {C.~G.}\ \bibnamefont {Sheu}},\ and\ \bibinfo {author}
  {\bibfnamefont {H.~R.}\ \bibnamefont {Weller}},\ }\bibfield  {title}
  {\bibinfo {title} {Energy levels of light nuclei {$A = 8$, $9$, $10$}},\
  }\href {https://doi.org/10.1016/j.nuclphysa.2004.09.059} {\bibfield
  {journal} {\bibinfo  {journal} {Nucl. Phys. A}\ }\textbf {\bibinfo {volume}
  {745}},\ \bibinfo {pages} {155} (\bibinfo {year} {2004})}\BibitemShut
  {NoStop}\bibitem [{\citenamefont {Kelley}\ \emph {et~al.}(2012)\citenamefont {Kelley},
  \citenamefont {Kwan}, \citenamefont {Purcell}, \citenamefont {Sheu},\ and\
  \citenamefont {Weller}}]{npa2012:011}\BibitemOpen
  \bibfield  {author} {\bibinfo {author} {\bibfnamefont {J.}~\bibnamefont
  {Kelley}}, \bibinfo {author} {\bibfnamefont {E.}~\bibnamefont {Kwan}},
  \bibinfo {author} {\bibfnamefont {J.~E.}\ \bibnamefont {Purcell}}, \bibinfo
  {author} {\bibfnamefont {C.~G.}\ \bibnamefont {Sheu}},\ and\ \bibinfo
  {author} {\bibfnamefont {H.~R.}\ \bibnamefont {Weller}},\ }\bibfield  {title}
  {\bibinfo {title} {Energy levels of light nuclei {$A = 11$}},\ }\href
  {https://doi.org/10.1016/j.nuclphysa.2012.01.010} {\bibfield  {journal}
  {\bibinfo  {journal} {Nucl. Phys. A}\ }\textbf {\bibinfo {volume} {880}},\
  \bibinfo {pages} {88} (\bibinfo {year} {2012})}\BibitemShut {NoStop}\bibitem [{\citenamefont {Kelley}\ \emph {et~al.}(2017)\citenamefont {Kelley},
  \citenamefont {Purcell},\ and\ \citenamefont {Sheu}}]{npa2017:012}\BibitemOpen
  \bibfield  {author} {\bibinfo {author} {\bibfnamefont {J.}~\bibnamefont
  {Kelley}}, \bibinfo {author} {\bibfnamefont {J.~E.}\ \bibnamefont
  {Purcell}},\ and\ \bibinfo {author} {\bibfnamefont {C.~G.}\ \bibnamefont
  {Sheu}},\ }\bibfield  {title} {\bibinfo {title} {Energy levels of light
  nuclei {$A = 12$}},\ }\href {https://doi.org/10.1016/j.nuclphysa.2017.07.015}
  {\bibfield  {journal} {\bibinfo  {journal} {Nucl. Phys. A}\ }\textbf
  {\bibinfo {volume} {968}},\ \bibinfo {pages} {71} (\bibinfo {year}
  {2017})}\BibitemShut {NoStop}\bibitem [{\citenamefont {Ajzenberg-Selove}(1991)}]{npa1991:013-015}\BibitemOpen
  \bibfield  {author} {\bibinfo {author} {\bibfnamefont {F.}~\bibnamefont
  {Ajzenberg-Selove}},\ }\bibfield  {title} {\bibinfo {title} {Energy levels of
  light nuclei {$A = 13$--$15$}},\ }\href
  {https://doi.org/10.1016/0375-9474(91)90446-D} {\bibfield  {journal}
  {\bibinfo  {journal} {Nucl. Phys. A}\ }\textbf {\bibinfo {volume} {523}},\
  \bibinfo {pages} {1} (\bibinfo {year} {1991})}\BibitemShut {NoStop}\bibitem [{\citenamefont {Neugart}\ and\ \citenamefont
  {Neyens}(2006)}]{neugart2006:nuclear-moments}\BibitemOpen
  \bibfield  {author} {\bibinfo {author} {\bibfnamefont {R.}~\bibnamefont
  {Neugart}}\ and\ \bibinfo {author} {\bibfnamefont {G.}~\bibnamefont
  {Neyens}},\ }\bibfield  {title} {\bibinfo {title} {Nuclear moments},\ }in\
  \href {https://doi.org/10.1007/3-540-33787-3_4} {\emph {\bibinfo {booktitle}
  {The Euroschool Lectures on Physics with Exotic Beams, Vol. II}}},\ \bibinfo
  {series} {Lecture Notes in Physics}, Vol.\ \bibinfo {volume} {700},\ \bibinfo
  {editor} {edited by\ \bibinfo {editor} {\bibfnamefont {J.}~\bibnamefont
  {Al-Khalili}}\ and\ \bibinfo {editor} {\bibfnamefont {E.}~\bibnamefont
  {Roeckl}}}\ (\bibinfo  {publisher} {Springer-Verlag},\ \bibinfo {address}
  {Berlin},\ \bibinfo {year} {2006})\ p.\ \bibinfo {pages} {135}\BibitemShut
  {NoStop}\bibitem [{\citenamefont {Brown}\ \emph {et~al.}(1991)\citenamefont {Brown},
  \citenamefont {Becchetti}, \citenamefont {J\"anecke}, \citenamefont
  {Ashktorab}, \citenamefont {Roberts}, \citenamefont {Kolata}, \citenamefont
  {Smith}, \citenamefont {Lamkin},\ and\ \citenamefont
  {Warner}}]{brown1991:8li-coulex}\BibitemOpen
  \bibfield  {author} {\bibinfo {author} {\bibfnamefont {J.~A.}\ \bibnamefont
  {Brown}}, \bibinfo {author} {\bibfnamefont {F.~D.}\ \bibnamefont
  {Becchetti}}, \bibinfo {author} {\bibfnamefont {J.~W.}\ \bibnamefont
  {J\"anecke}}, \bibinfo {author} {\bibfnamefont {K.}~\bibnamefont
  {Ashktorab}}, \bibinfo {author} {\bibfnamefont {D.~A.}\ \bibnamefont
  {Roberts}}, \bibinfo {author} {\bibfnamefont {J.~J.}\ \bibnamefont {Kolata}},
  \bibinfo {author} {\bibfnamefont {R.~J.}\ \bibnamefont {Smith}}, \bibinfo
  {author} {\bibfnamefont {K.}~\bibnamefont {Lamkin}},\ and\ \bibinfo {author}
  {\bibfnamefont {R.~E.}\ \bibnamefont {Warner}},\ }\bibfield  {title}
  {\bibinfo {title} {Coulomb excitation of $\isotope[8]{Li}$},\ }\href
  {https://doi.org/10.1103/PhysRevLett.66.2452} {\bibfield  {journal} {\bibinfo
   {journal} {Phys. Rev. Lett.}\ }\textbf {\bibinfo {volume} {66}},\ \bibinfo
  {pages} {2452} (\bibinfo {year} {1991})}\BibitemShut {NoStop}\bibitem [{\citenamefont {Weisskopf}(1951)}]{weisskopf1951:estimate}\BibitemOpen
  \bibfield  {author} {\bibinfo {author} {\bibfnamefont {V.~F.}\ \bibnamefont
  {Weisskopf}},\ }\bibfield  {title} {\bibinfo {title} {Radiative transition
  probabilities in nuclei},\ }\href {https://doi.org/10.1103/PhysRev.83.1073}
  {\bibfield  {journal} {\bibinfo  {journal} {Phys. Rev.}\ }\textbf {\bibinfo
  {volume} {83}},\ \bibinfo {pages} {1073} (\bibinfo {year}
  {1951})}\BibitemShut {NoStop}\bibitem [{\citenamefont {Pastore}\ \emph {et~al.}(2013)\citenamefont
  {Pastore}, \citenamefont {Pieper}, \citenamefont {Schiavilla},\ and\
  \citenamefont {Wiringa}}]{pastore2013:qmc-em-alt9}\BibitemOpen
  \bibfield  {author} {\bibinfo {author} {\bibfnamefont {S.}~\bibnamefont
  {Pastore}}, \bibinfo {author} {\bibfnamefont {S.~C.}\ \bibnamefont {Pieper}},
  \bibinfo {author} {\bibfnamefont {R.}~\bibnamefont {Schiavilla}},\ and\
  \bibinfo {author} {\bibfnamefont {R.~B.}\ \bibnamefont {Wiringa}},\
  }\bibfield  {title} {\bibinfo {title} {Quantum {M}onte {C}arlo calculations
  of electromagnetic moments and transitions in {$A\le9$} nuclei with
  meson-exchange currents derived from chiral effective field theory},\ }\href
  {https://doi.org/10.1103/PhysRevC.87.035503} {\bibfield  {journal} {\bibinfo
  {journal} {Phys. Rev. C}\ }\textbf {\bibinfo {volume} {87}},\ \bibinfo
  {pages} {035503} (\bibinfo {year} {2013})}\BibitemShut {NoStop}\bibitem [{\citenamefont {Suhonen}(2007)}]{suhonen2007:nucleons-nucleus}\BibitemOpen
  \bibfield  {author} {\bibinfo {author} {\bibfnamefont {J.}~\bibnamefont
  {Suhonen}},\ }\href {https://doi.org/10.1007/978-3-540-48861-3} {\emph
  {\bibinfo {title} {From Nucleons to Nucleus}}}\ (\bibinfo  {publisher}
  {Springer-Verlag},\ \bibinfo {address} {Berlin},\ \bibinfo {year}
  {2007})\BibitemShut {NoStop}\bibitem [{\citenamefont {Shirokov}\ \emph {et~al.}(2016)\citenamefont
  {Shirokov}, \citenamefont {Shin}, \citenamefont {Kim}, \citenamefont
  {Sosonkina}, \citenamefont {Maris},\ and\ \citenamefont
  {Vary}}]{shirokov2016:nn-daejeon16}\BibitemOpen
  \bibfield  {author} {\bibinfo {author} {\bibfnamefont {A.~M.}\ \bibnamefont
  {Shirokov}}, \bibinfo {author} {\bibfnamefont {I.~J.}\ \bibnamefont {Shin}},
  \bibinfo {author} {\bibfnamefont {Y.}~\bibnamefont {Kim}}, \bibinfo {author}
  {\bibfnamefont {M.}~\bibnamefont {Sosonkina}}, \bibinfo {author}
  {\bibfnamefont {P.}~\bibnamefont {Maris}},\ and\ \bibinfo {author}
  {\bibfnamefont {J.~P.}\ \bibnamefont {Vary}},\ }\bibfield  {title} {\bibinfo
  {title} {{N3LO} {$NN$} interaction adjusted to light nuclei in \textit{ab
  exitu} approach},\ }\href {https://doi.org/10.1016/j.physletb.2016.08.006}
  {\bibfield  {journal} {\bibinfo  {journal} {Phys. Lett. B}\ }\textbf
  {\bibinfo {volume} {761}},\ \bibinfo {pages} {87} (\bibinfo {year}
  {2016})}\BibitemShut {NoStop}\bibitem [{\citenamefont {Entem}\ and\ \citenamefont
  {Machleidt}(2003)}]{entem2003:chiral-nn-potl}\BibitemOpen
  \bibfield  {author} {\bibinfo {author} {\bibfnamefont {D.~R.}\ \bibnamefont
  {Entem}}\ and\ \bibinfo {author} {\bibfnamefont {R.}~\bibnamefont
  {Machleidt}},\ }\bibfield  {title} {\bibinfo {title} {Accurate
  charge-dependent nucleon-nucleon potential at fourth order of chiral
  perturbation theory},\ }\href {https://doi.org/10.1103/PhysRevC.68.041001}
  {\bibfield  {journal} {\bibinfo  {journal} {Phys. Rev. C}\ }\textbf {\bibinfo
  {volume} {68}},\ \bibinfo {pages} {041001(R)} (\bibinfo {year}
  {2003})}\BibitemShut {NoStop}\bibitem [{\citenamefont {Bogner}\ \emph {et~al.}(2007)\citenamefont {Bogner},
  \citenamefont {Furnstahl},\ and\ \citenamefont
  {Perry}}]{bogner2007:srg-nucleon}\BibitemOpen
  \bibfield  {author} {\bibinfo {author} {\bibfnamefont {S.~K.}\ \bibnamefont
  {Bogner}}, \bibinfo {author} {\bibfnamefont {R.~J.}\ \bibnamefont
  {Furnstahl}},\ and\ \bibinfo {author} {\bibfnamefont {R.~J.}\ \bibnamefont
  {Perry}},\ }\bibfield  {title} {\bibinfo {title} {Similarity renormalization
  group for nucleon-nucleon interactions},\ }\href
  {https://doi.org/10.1103/PhysRevC.75.061001} {\bibfield  {journal} {\bibinfo
  {journal} {Phys. Rev. C}\ }\textbf {\bibinfo {volume} {75}},\ \bibinfo
  {pages} {061001(R)} (\bibinfo {year} {2007})}\BibitemShut {NoStop}\bibitem [{\citenamefont {Aktulga}\ \emph {et~al.}(2014)\citenamefont
  {Aktulga}, \citenamefont {Yang}, \citenamefont {Ng}, \citenamefont {Maris},\
  and\ \citenamefont {Vary}}]{aktulga2014:mfdn-scalability}\BibitemOpen
  \bibfield  {author} {\bibinfo {author} {\bibfnamefont {H.~M.}\ \bibnamefont
  {Aktulga}}, \bibinfo {author} {\bibfnamefont {C.}~\bibnamefont {Yang}},
  \bibinfo {author} {\bibfnamefont {E.~G.}\ \bibnamefont {Ng}}, \bibinfo
  {author} {\bibfnamefont {P.}~\bibnamefont {Maris}},\ and\ \bibinfo {author}
  {\bibfnamefont {J.~P.}\ \bibnamefont {Vary}},\ }\bibfield  {title} {\bibinfo
  {title} {Improving the scalability of symmetric iterative eigensolver for
  multi-core platforms},\ }\href {https://doi.org/10.1002/cpe.3129} {\bibfield
  {journal} {\bibinfo  {journal} {Concurrency Computat.: Pract. Exper.}\
  }\textbf {\bibinfo {volume} {26}},\ \bibinfo {pages} {2631} (\bibinfo {year}
  {2014})}\BibitemShut {NoStop}\bibitem [{\citenamefont {Shao}\ \emph {et~al.}(2018)\citenamefont {Shao},
  \citenamefont {Aktulga}, \citenamefont {Yang}, \citenamefont {Ng},
  \citenamefont {Maris},\ and\ \citenamefont
  {Vary}}]{shao2018:ncci-preconditioned}\BibitemOpen
  \bibfield  {author} {\bibinfo {author} {\bibfnamefont {M.}~\bibnamefont
  {Shao}}, \bibinfo {author} {\bibfnamefont {H.~M.}\ \bibnamefont {Aktulga}},
  \bibinfo {author} {\bibfnamefont {C.}~\bibnamefont {Yang}}, \bibinfo {author}
  {\bibfnamefont {E.~G.}\ \bibnamefont {Ng}}, \bibinfo {author} {\bibfnamefont
  {P.}~\bibnamefont {Maris}},\ and\ \bibinfo {author} {\bibfnamefont {J.~P.}\
  \bibnamefont {Vary}},\ }\bibfield  {title} {\bibinfo {title} {Accelerating
  nuclear configuration interaction calculations through a preconditioned block
  iterative eigensolver},\ }\href {https://doi.org/10.1016/j.cpc.2017.09.004}
  {\bibfield  {journal} {\bibinfo  {journal} {Comput. Phys. Commun.}\ }\textbf
  {\bibinfo {volume} {222}},\ \bibinfo {pages} {1} (\bibinfo {year}
  {2018})}\BibitemShut {NoStop}\bibitem [{\citenamefont {Cook}\ \emph {et~al.}(2022)\citenamefont {Cook},
  \citenamefont {Fasano}, \citenamefont {Maris}, \citenamefont {Yang},\ and\
  \citenamefont {Oryspayev}}]{cook2022:mfdn-directives-waccpd21}\BibitemOpen
  \bibfield  {author} {\bibinfo {author} {\bibfnamefont {B.~G.}\ \bibnamefont
  {Cook}}, \bibinfo {author} {\bibfnamefont {P.~J.}\ \bibnamefont {Fasano}},
  \bibinfo {author} {\bibfnamefont {P.}~\bibnamefont {Maris}}, \bibinfo
  {author} {\bibfnamefont {C.}~\bibnamefont {Yang}},\ and\ \bibinfo {author}
  {\bibfnamefont {D.}~\bibnamefont {Oryspayev}},\ }\bibfield  {title} {\bibinfo
  {title} {Accelerating quantum many-body configuration interaction with
  directives},\ }in\ \href {https://doi.org/10.1007/978-3-030-97759-7_6} {\emph
  {\bibinfo {booktitle} {Accelerator Programming Using Directives}}},\ \bibinfo
  {series} {Lect. Notes Comput. Sci.}, Vol.\ \bibinfo {volume} {13194},\
  \bibinfo {editor} {edited by\ \bibinfo {editor} {\bibfnamefont
  {S.}~\bibnamefont {Bhalachandra}}, \bibinfo {editor} {\bibfnamefont
  {C.}~\bibnamefont {Daley}},\ and\ \bibinfo {editor} {\bibfnamefont
  {V.}~\bibnamefont {Melesse~Vergara}}}\ (\bibinfo  {publisher}
  {Springer-Verlag},\ \bibinfo {address} {Berlin},\ \bibinfo {year} {2022})\
  pp.\ \bibinfo {pages} {112--132}\BibitemShut {NoStop}\bibitem [{sup()}]{supplemental-material}\BibitemOpen
  \bibinfo {note} {See Supplemental Material at http://xxxxxxxx for
  comprehensive plots and tabulations of the calculated observables, as
  functions of $\Nmax$ and $\hw$.}\BibitemShut {Stop}\bibitem [{\citenamefont {Carlson}\ \emph {et~al.}(2015)\citenamefont
  {Carlson}, \citenamefont {Gandolfi}, \citenamefont {Pederiva}, \citenamefont
  {Pieper}, \citenamefont {Schiavilla}, \citenamefont {Schmidt},\ and\
  \citenamefont {Wiringa}}]{carlson2015:qmc-nuclear}\BibitemOpen
  \bibfield  {author} {\bibinfo {author} {\bibfnamefont {J.}~\bibnamefont
  {Carlson}}, \bibinfo {author} {\bibfnamefont {S.}~\bibnamefont {Gandolfi}},
  \bibinfo {author} {\bibfnamefont {F.}~\bibnamefont {Pederiva}}, \bibinfo
  {author} {\bibfnamefont {S.~C.}\ \bibnamefont {Pieper}}, \bibinfo {author}
  {\bibfnamefont {R.}~\bibnamefont {Schiavilla}}, \bibinfo {author}
  {\bibfnamefont {K.~E.}\ \bibnamefont {Schmidt}},\ and\ \bibinfo {author}
  {\bibfnamefont {R.~B.}\ \bibnamefont {Wiringa}},\ }\bibfield  {title}
  {\bibinfo {title} {Quantum {M}onte {C}arlo methods for nuclear physics},\
  }\href {https://doi.org/10.1103/RevModPhys.87.1067} {\bibfield  {journal}
  {\bibinfo  {journal} {Rev. Mod. Phys.}\ }\textbf {\bibinfo {volume} {87}},\
  \bibinfo {pages} {1067} (\bibinfo {year} {2015})}\BibitemShut {NoStop}\bibitem [{\citenamefont {Millener}(2001)}]{millener2001:light-nuclei}\BibitemOpen
  \bibfield  {author} {\bibinfo {author} {\bibfnamefont {D.~J.}\ \bibnamefont
  {Millener}},\ }\bibfield  {title} {\bibinfo {title} {Structure of unstable
  light nuclei},\ }\href {https://doi.org/10.1016/S0375-9474(01)00589-9}
  {\bibfield  {journal} {\bibinfo  {journal} {Nucl. Phys. A}\ }\textbf
  {\bibinfo {volume} {693}},\ \bibinfo {pages} {394} (\bibinfo {year}
  {2001})}\BibitemShut {NoStop}\bibitem [{\citenamefont {Millener}(2007)}]{millener2007:p-shell-hypernuclei}\BibitemOpen
  \bibfield  {author} {\bibinfo {author} {\bibfnamefont {D.~J.}\ \bibnamefont
  {Millener}},\ }\bibinfo {title} {Hypernuclear gamma-ray spectroscopy and the
  structure of $p$-shell nuclei and hypernuclei},\ in\ \href
  {https://doi.org/10.1007/978-3-540-72039-3_2} {\emph {\bibinfo {booktitle}
  {Topics in Strangeness Nuclear Physics}}},\ \bibinfo {series} {Lecture Notes
  in Physics}, Vol.\ \bibinfo {volume} {724},\ \bibinfo {editor} {edited by\
  \bibinfo {editor} {\bibfnamefont {P.}~\bibnamefont {Byd{\v{z}}ovsk{\'y}}},
  \bibinfo {editor} {\bibfnamefont {J.}~\bibnamefont {Mare{\v{s}}}},\ and\
  \bibinfo {editor} {\bibfnamefont {A.}~\bibnamefont {Gal}}}\ (\bibinfo
  {publisher} {Springer},\ \bibinfo {address} {Berlin},\ \bibinfo {year}
  {2007})\ p.~\bibinfo {pages} {31}\BibitemShut {NoStop}\bibitem [{\citenamefont {Shirokov}\ \emph {et~al.}(2007)\citenamefont
  {Shirokov}, \citenamefont {Vary}, \citenamefont {Mazur},\ and\ \citenamefont
  {Weber}}]{shirokov2007:nn-jisp16}\BibitemOpen
  \bibfield  {author} {\bibinfo {author} {\bibfnamefont {A.~M.}\ \bibnamefont
  {Shirokov}}, \bibinfo {author} {\bibfnamefont {J.~P.}\ \bibnamefont {Vary}},
  \bibinfo {author} {\bibfnamefont {A.~I.}\ \bibnamefont {Mazur}},\ and\
  \bibinfo {author} {\bibfnamefont {T.~A.}\ \bibnamefont {Weber}},\ }\bibfield
  {title} {\bibinfo {title} {Realistic nuclear {H}amiltonian: \textit{Ab exitu}
  approach},\ }\href {https://doi.org/10.1016/j.physletb.2006.10.066}
  {\bibfield  {journal} {\bibinfo  {journal} {Phys. Lett. B}\ }\textbf
  {\bibinfo {volume} {644}},\ \bibinfo {pages} {33} (\bibinfo {year}
  {2007})}\BibitemShut {NoStop}\bibitem [{\citenamefont {Epelbaum}\ \emph
  {et~al.}(2015{\natexlab{a}})\citenamefont {Epelbaum}, \citenamefont {Krebs},\
  and\ \citenamefont {Mei{\ss}ner}}]{epelbaum2015:lenpic-n4lo-scs}\BibitemOpen
  \bibfield  {author} {\bibinfo {author} {\bibfnamefont {E.}~\bibnamefont
  {Epelbaum}}, \bibinfo {author} {\bibfnamefont {H.}~\bibnamefont {Krebs}},\
  and\ \bibinfo {author} {\bibfnamefont {U.-G.}\ \bibnamefont {Mei{\ss}ner}},\
  }\bibfield  {title} {\bibinfo {title} {Precision nucleon-nucleon potential at
  fifth order in the chiral expansion},\ }\href
  {https://doi.org/10.1103/PhysRevLett.115.122301} {\bibfield  {journal}
  {\bibinfo  {journal} {Phys. Rev. Lett.}\ }\textbf {\bibinfo {volume} {115}},\
  \bibinfo {pages} {122301} (\bibinfo {year} {2015}{\natexlab{a}})}\BibitemShut
  {NoStop}\bibitem [{\citenamefont {Epelbaum}\ \emph
  {et~al.}(2015{\natexlab{b}})\citenamefont {Epelbaum}, \citenamefont {Krebs},\
  and\ \citenamefont {Mei{\ss}ner}}]{epelbaum2015:lenpic-n3lo-scs}\BibitemOpen
  \bibfield  {author} {\bibinfo {author} {\bibfnamefont {E.}~\bibnamefont
  {Epelbaum}}, \bibinfo {author} {\bibfnamefont {H.}~\bibnamefont {Krebs}},\
  and\ \bibinfo {author} {\bibfnamefont {U.-G.}\ \bibnamefont {Mei{\ss}ner}},\
  }\bibfield  {title} {\bibinfo {title} {Improved chiral nucleon-nucleon
  potential up to next-to-next-to-next-to-leading order},\ }\href
  {https://doi.org/10.1140/epja/i2015-15053-8} {\bibfield  {journal} {\bibinfo
  {journal} {Eur. Phys. J. A}\ }\textbf {\bibinfo {volume} {51}},\ \bibinfo
  {pages} {53} (\bibinfo {year} {2015}{\natexlab{b}})}\BibitemShut {NoStop}\bibitem [{\citenamefont {Wiringa}\ \emph {et~al.}(1995)\citenamefont
  {Wiringa}, \citenamefont {Stoks},\ and\ \citenamefont
  {Schiavilla}}]{wiringa1995:nn-av18}\BibitemOpen
  \bibfield  {author} {\bibinfo {author} {\bibfnamefont {R.~B.}\ \bibnamefont
  {Wiringa}}, \bibinfo {author} {\bibfnamefont {V.~G.~J.}\ \bibnamefont
  {Stoks}},\ and\ \bibinfo {author} {\bibfnamefont {R.}~\bibnamefont
  {Schiavilla}},\ }\bibfield  {title} {\bibinfo {title} {Accurate
  nucleon-nucleon potential with charge-independence breaking},\ }\href
  {https://doi.org/10.1103/PhysRevC.51.38} {\bibfield  {journal} {\bibinfo
  {journal} {Phys. Rev. C}\ }\textbf {\bibinfo {volume} {51}},\ \bibinfo
  {pages} {38} (\bibinfo {year} {1995})}\BibitemShut {NoStop}\bibitem [{\citenamefont {Pieper}\ \emph {et~al.}(2001)\citenamefont {Pieper},
  \citenamefont {Pandharipande}, \citenamefont {Wiringa},\ and\ \citenamefont
  {Carlson}}]{pieper2001:3n-il2}\BibitemOpen
  \bibfield  {author} {\bibinfo {author} {\bibfnamefont {S.~C.}\ \bibnamefont
  {Pieper}}, \bibinfo {author} {\bibfnamefont {V.~R.}\ \bibnamefont
  {Pandharipande}}, \bibinfo {author} {\bibfnamefont {R.~B.}\ \bibnamefont
  {Wiringa}},\ and\ \bibinfo {author} {\bibfnamefont {J.}~\bibnamefont
  {Carlson}},\ }\bibfield  {title} {\bibinfo {title} {Realistic models of
  pion-exchange three-nucleon interactions},\ }\href
  {https://doi.org/10.1103/PhysRevC.64.014001} {\bibfield  {journal} {\bibinfo
  {journal} {Phys. Rev. C}\ }\textbf {\bibinfo {volume} {64}},\ \bibinfo
  {pages} {014001} (\bibinfo {year} {2001})}\BibitemShut {NoStop}\bibitem [{\citenamefont {Lane}(1960)}]{lane1960:reduced-widths}\BibitemOpen
  \bibfield  {author} {\bibinfo {author} {\bibfnamefont {A.~M.}\ \bibnamefont
  {Lane}},\ }\bibfield  {title} {\bibinfo {title} {Reduced widths of individual
  nuclear energy levels},\ }\href {https://doi.org/10.1103/RevModPhys.32.519}
  {\bibfield  {journal} {\bibinfo  {journal} {Rev. Mod. Phys.}\ }\textbf
  {\bibinfo {volume} {32}},\ \bibinfo {pages} {519} (\bibinfo {year}
  {1960})}\BibitemShut {NoStop}\bibitem [{\citenamefont {Maris}\ \emph {et~al.}(2013)\citenamefont {Maris},
  \citenamefont {Vary},\ and\ \citenamefont
  {Navr{\'a}til}}]{maris2013:ncsm-chiral-a7-a8}\BibitemOpen
  \bibfield  {author} {\bibinfo {author} {\bibfnamefont {P.}~\bibnamefont
  {Maris}}, \bibinfo {author} {\bibfnamefont {J.~P.}\ \bibnamefont {Vary}},\
  and\ \bibinfo {author} {\bibfnamefont {P.}~\bibnamefont {Navr{\'a}til}},\
  }\bibfield  {title} {\bibinfo {title} {Structure of {$A=7$}--$8$ nuclei with
  two- plus three-nucleon interactions from chiral effective field theory},\
  }\href {https://doi.org/10.1103/PhysRevC.87.014327} {\bibfield  {journal}
  {\bibinfo  {journal} {Phys. Rev. C}\ }\textbf {\bibinfo {volume} {87}},\
  \bibinfo {pages} {014327} (\bibinfo {year} {2013})}\BibitemShut {NoStop}\bibitem [{\citenamefont {Navr{\'a}til}(2007)}]{navratil2007:local-chiral-3n}\BibitemOpen
  \bibfield  {author} {\bibinfo {author} {\bibfnamefont {P.}~\bibnamefont
  {Navr{\'a}til}},\ }\bibfield  {title} {\bibinfo {title} {Local three-nucleon
  interaction from chiral effective field theory},\ }\href
  {https://doi.org/10.1007/s00601-007-0193-3} {\bibfield  {journal} {\bibinfo
  {journal} {Few-Body Syst.}\ }\textbf {\bibinfo {volume} {41}},\ \bibinfo
  {pages} {117} (\bibinfo {year} {2007})}\BibitemShut {NoStop}\bibitem [{\citenamefont
  {{\^O}kubo}(1954)}]{okubo1954:diagonalization-tamm-dancoff}\BibitemOpen
  \bibfield  {author} {\bibinfo {author} {\bibfnamefont {S.}~\bibnamefont
  {{\^O}kubo}},\ }\bibfield  {title} {\bibinfo {title} {Diagonalization of
  {Hamiltonian} and {Tamm-Dancoff} equation},\ }\href
  {https://doi.org/10.1143/PTP.12.603} {\bibfield  {journal} {\bibinfo
  {journal} {Prog. Theor. Phys.}\ }\textbf {\bibinfo {volume} {12}},\ \bibinfo
  {pages} {603} (\bibinfo {year} {1954})}\BibitemShut {NoStop}\bibitem [{\citenamefont {Suzuki}\ and\ \citenamefont
  {Lee}(1980)}]{suzuki1980:effective-interaction}\BibitemOpen
  \bibfield  {author} {\bibinfo {author} {\bibfnamefont {K.}~\bibnamefont
  {Suzuki}}\ and\ \bibinfo {author} {\bibfnamefont {S.~Y.}\ \bibnamefont
  {Lee}},\ }\bibfield  {title} {\bibinfo {title} {Convergent theory for
  effective interaction in nuclei},\ }\href
  {https://doi.org/10.1143/PTP.64.2091} {\bibfield  {journal} {\bibinfo
  {journal} {Prog. Theor. Phys.}\ }\textbf {\bibinfo {volume} {64}},\ \bibinfo
  {pages} {2091} (\bibinfo {year} {1980})}\BibitemShut {NoStop}\bibitem [{\citenamefont
  {Elliott}(1958{\natexlab{a}})}]{elliott1958:su3-part1}\BibitemOpen
  \bibfield  {author} {\bibinfo {author} {\bibfnamefont {J.~P.}\ \bibnamefont
  {Elliott}},\ }\bibfield  {title} {\bibinfo {title} {Collective motion in the
  nuclear shell model. {I}. {C}lassification schemes for states of mixed
  configurations},\ }\href {https://doi.org/10.1098/rspa.1958.0072} {\bibfield
  {journal} {\bibinfo  {journal} {Proc. R. Soc. London A}\ }\textbf {\bibinfo
  {volume} {245}},\ \bibinfo {pages} {128} (\bibinfo {year}
  {1958}{\natexlab{a}})}\BibitemShut {NoStop}\bibitem [{\citenamefont
  {Elliott}(1958{\natexlab{b}})}]{elliott1958:su3-part2}\BibitemOpen
  \bibfield  {author} {\bibinfo {author} {\bibfnamefont {J.~P.}\ \bibnamefont
  {Elliott}},\ }\bibfield  {title} {\bibinfo {title} {Collective motion in the
  nuclear shell model. {II}. {T}he introduction of intrinsic wave functions},\
  }\href {https://doi.org/10.1098/rspa.1958.0101} {\bibfield  {journal}
  {\bibinfo  {journal} {Proc. R. Soc. London A}\ }\textbf {\bibinfo {volume}
  {245}},\ \bibinfo {pages} {562} (\bibinfo {year}
  {1958}{\natexlab{b}})}\BibitemShut {NoStop}\bibitem [{\citenamefont {Elliott}\ and\ \citenamefont
  {Harvey}(1963)}]{elliott1963:su3-part3}\BibitemOpen
  \bibfield  {author} {\bibinfo {author} {\bibfnamefont {J.~P.}\ \bibnamefont
  {Elliott}}\ and\ \bibinfo {author} {\bibfnamefont {M.}~\bibnamefont
  {Harvey}},\ }\bibfield  {title} {\bibinfo {title} {Collective motion in the
  nuclear shell model. {III}. {T}he calculation of spectra},\ }\href
  {https://doi.org/10.1098/rspa.1963.0071} {\bibfield  {journal} {\bibinfo
  {journal} {Proc. R. Soc. London A}\ }\textbf {\bibinfo {volume} {272}},\
  \bibinfo {pages} {557} (\bibinfo {year} {1963})}\BibitemShut {NoStop}\bibitem [{\citenamefont {Elliott}\ and\ \citenamefont
  {Wilsdon}(1968)}]{elliott1968:su3-part4}\BibitemOpen
  \bibfield  {author} {\bibinfo {author} {\bibfnamefont {J.~P.}\ \bibnamefont
  {Elliott}}\ and\ \bibinfo {author} {\bibfnamefont {C.~E.}\ \bibnamefont
  {Wilsdon}},\ }\bibfield  {title} {\bibinfo {title} {Collective motion in the
  nuclear shell model. {IV}. {O}dd-mass nuclei in the $sd$ shell},\ }\href
  {https://doi.org/10.1098/rspa.1968.0033} {\bibfield  {journal} {\bibinfo
  {journal} {Proc. R. Soc. London A}\ }\textbf {\bibinfo {volume} {302}},\
  \bibinfo {pages} {509} (\bibinfo {year} {1968})}\BibitemShut {NoStop}\bibitem [{\citenamefont {Harvey}(1968)}]{harvey1968:su3-shell}\BibitemOpen
  \bibfield  {author} {\bibinfo {author} {\bibfnamefont {M.}~\bibnamefont
  {Harvey}},\ }\bibinfo {title} {The nuclear $\mathit{SU}_3$ model},\ in\ \href
  {https://doi.org/10.1007/978-1-4757-0103-6_2} {\emph {\bibinfo {booktitle}
  {Advances in Nuclear Physics}}},\ Vol.~\bibinfo {volume} {1},\ \bibinfo
  {editor} {edited by\ \bibinfo {editor} {\bibfnamefont {M.}~\bibnamefont
  {Baranger}}\ and\ \bibinfo {editor} {\bibfnamefont {E.}~\bibnamefont
  {Vogt}}}\ (\bibinfo  {publisher} {Plenum},\ \bibinfo {address} {New York},\
  \bibinfo {year} {1968})\ p.~\bibinfo {pages} {67}\BibitemShut {NoStop}\bibitem [{\citenamefont {Heil}\ \emph {et~al.}(1998)\citenamefont {Heil},
  \citenamefont {Kappeler}, \citenamefont {Wiescher},\ and\ \citenamefont
  {Mengoni}}]{heil1998:7li-ngamma}\BibitemOpen
  \bibfield  {author} {\bibinfo {author} {\bibfnamefont {M.}~\bibnamefont
  {Heil}}, \bibinfo {author} {\bibfnamefont {F.}~\bibnamefont {Kappeler}},
  \bibinfo {author} {\bibfnamefont {M.}~\bibnamefont {Wiescher}},\ and\
  \bibinfo {author} {\bibfnamefont {A.}~\bibnamefont {Mengoni}},\ }\bibfield
  {title} {\bibinfo {title} {The $(n, \gamma)$ cross section of
  $\isotope[7]{Li}$},\ }\href {https://doi.org/10.1086/306367} {\bibfield
  {journal} {\bibinfo  {journal} {Astrophys. J.}\ }\textbf {\bibinfo {volume}
  {507}},\ \bibinfo {pages} {997} (\bibinfo {year} {1998})}\BibitemShut
  {NoStop}\end{thebibliography}
\nocite{control:title-on}

\end{document}